\documentclass[11pt,a4paper]{article}
\pdfoutput=1

\usepackage{jcappub}
\usepackage{amsmath}

\usepackage{multirow}

\newcommand{\Haslam}{$408\,\text{MHz}$}
\newcommand{\HaslamASS}{$408\,\text{MHz}$ all-sky survey}

\newcommand{\dd}{\mathrm{d}}
\newcommand{\ee}{\mathrm{e}}
\renewcommand{\vec}[1]{{\bf #1}}

\title{Loops and spurs: The angular power spectrum of the Galactic synchrotron background}

\author[a]{Philipp Mertsch} \author[b,c]{and Subir Sarkar}
\affiliation[a]{Kavli Institute for Particle Astrophysics \&
  Cosmology, 2575 Sand Hill Road, M/S 29, Menlo Park, CA 94025, USA}
\affiliation[b]{(On sabbatical leave from) Rudolf Peierls Centre for
  Theoretical Physics, University of Oxford, 1 Keble Road, Oxford OX1
  3NP, UK} 
\affiliation[c]{(Visiting Professor at) Niels Bohr Institute,
  Copenhagen University, Blegdamsvej 17, 2100 Copenhagen \O, Denmark}
\emailAdd{pmertsch@stanford.edu, sarkar@nbi.dk}

\abstract{We present a new model of the diffuse Galactic synchrotron
  radiation, concentrating on its angular anisotropies. While previous
  studies have focussed on either the variation of the emissivity on
  large ($\sim$~kpc) scales, or on fluctuations due to MHD turbulence
  in the interstellar medium, we unify these approaches to match the
  angular power spectrum. We note that the usual turbulence cascade
  calculation ignores spatial correlations at the injection scale
  ($\sim100$~pc) due to compression of the interstellar medium by of
  ${\cal O}(1000)$ old supernova remnants --- the `radio loops' only
  four of which are visible in radio maps. This new component
  naturally provides the otherwise missing power on intermediate and
  small scales in the all-sky map at 408 MHz. Our model can enable
  more reliable subtraction of the synchrotron foreground for studies
  of CMB anisotropies or searches for dark matter annihilation. We
  conclude with some remarks on the relevance to modelling of the
  polarised foreground.}


\begin{document}
\maketitle
\flushbottom

\section{Introduction}

Radio maps of the sky contain a wealth of information on non-thermal
astrophysical processes. The sources of this radiation range from
compact objects like pulsars with magnetospheres extending over
hundreds of kilometers, to radio galaxies hundreds of kiloparsecs
across, thus spanning over 16 orders of magnitude in size. The total
brightness of the radio sky is dominated by diffuse emission from the
Galaxy at frequencies between tens of MHz and a few GHz
\cite{1964ARA&A...2..185M}. It is produced by synchrotron radiation of
cosmic ray (CR) electrons and positrons\footnote{In the following,
  when using the term `electrons' we refer to both electrons and
  positrons.} gyrating in the Galactic magnetic field (GMF) and for
typical field strengths of ${\cal O}(\mu$G), the CR electron spectrum
is probed between hundreds of MeV and tens of GeV. The frequency and
angular dependence of this diffuse emission encodes both the
properties of the CR electron sources, e.g. their spectrum and spatial
distribution, and of the GMF and its effect on the propagation of
CRs. Therefore, studies of the diffuse radio background can in
principle determine crucial parameters of the propagation of CRs,
e.g. the size of the (radio) halo in which they are confined.

The diffuse synchrotron emission at GHz frequencies also constitutes
an important foreground for studies of cosmologically important
angular anisotropies in the cosmic microwave background (CMB)
\cite{Hu:2001bc}. Only in a narrow window around 70~GHz does the CMB
anisotropy signal dominate over synchrotron, dust and thermal
bremsstrahlung emission from the Milky Way.  For example, with a
Galactic plane mask that preserves 77\% of the sky the RMS CMB
anisotropy begins to exceed the (falling) synchrotron foreground
around 30~GHz and drops below the (rising) foreground due to dust at
about 160~GHz, while it dominates by a factor of $\sim 8$ between 70
and 80~GHz --- see Fig.~10 of~\cite{Bennett:2003ca}. In particular,
detection of the elusive $B$-mode polarisation, a diagnostic of
gravitational waves expected to have been generated during primordial
inflation, requires detailed understanding and reliable subtraction of
the Galactic `foreground'
\cite{Dunkley:2008am,Bouchet:2011ck,Delabrouille:2012ye}. This is also
a limiting factor for the sensitivity of radio searches for dark
matter annihilation signals as has been amply illustrated in the
context of the `WMAP haze' \cite{Mertsch:2010ga,Planck:2012fb}.

As regards the GMF, it is customary to distinguish between an ordered
and a `random' component; the former is coherent on ${\cal O}$(kpc)
scales while the latter varies on scales of ${\cal O}(100)\,
\text{pc}$. Consequently, the specific synchrotron emissivity,
i.e. the power emitted per unit volume and frequency, varies on two
(disparate) scales. On small ($\lesssim 100 \, \text{pc}$) scales the
variation is almost entirely due to the turbulent nature of the
interstellar magnetic field. It is believed that from the `outer
scale' ($L \sim 100\, \text{pc}$) on which energy is injected into the
interstellar medium (ISM) --- most likely from old supernova remnant
(SNR) shock waves --- the turbulence cascades down to smaller
scales. Both theoretical
arguments~\cite{Kolmogorov:1941aa,Goldreich:1997aa} and MHD
simulations suggest a power law for the turbulent energy in magnetic
fields:
\begin{equation}
\dd B(k)^2/\dd^3 k \propto k^{-\alpha}, \quad \alpha =11/3 \, . 
\label{Kolmogorov}
\end{equation}
The spatial correlations of the magnetic field lead to angular
correlations of the synchrotron flux $J$, $C(\psi) \equiv \langle
J(\vec{n}) J(\vec{n}') \rangle$, where $\vec{n}$ and $\vec{n}'$ are
two directions on the sky and $\psi = \arccos{(\vec{n} \cdot
  \vec{n}')}$ is the angle between them. A convenient representation
for studying these correlations is the angular power spectrum (APS):
\begin{equation}
\mathcal{C}_l = \int \dd (\cos{\psi}) P_l(\cos{\psi}) C (\psi) \, , 
\label{eqn:defAPS1}
\end{equation}
i.e. the Legendre transform of the angular two-point correlation
function $C(\psi)$ --- structure on angular scales $\psi$ is encoded
in multipoles $l \sim \pi/\psi$. The APS of a sky map $J(\theta,
\phi)$ can also be calculated from its spherical harmonics
coefficients $a_{lm}$:
\begin{equation}
\mathcal{C}_l \equiv \frac{1}{2 l+1} \sum_{m=-l}^{l} \left| a_{lm} \right|^2
\quad \text{where} \quad
a_{lm} = \int \dd \Omega \, Y_{lm}^*(\theta, \phi) J(\theta, \phi) \, .
\label{eqn:defAPS2}
\end{equation}
This is equivalent to the definition in eq.~(\ref{eqn:defAPS1}) for a
(statistically) isotropic sky; here, $Y_{lm}(\theta, \phi)$ denotes
the spherical harmonics function.

For a power law spectrum of magnetic fluctuations (see
eq.\ref{Kolmogorov}), quasi-linear theory for wave-particle
interactions predicts a rigidity-dependent diffusion coefficient which
is also a power law with a related index, $D_{xx} \propto
\mathcal{R}^{\delta}$ with $\delta = 4-\alpha$ and a normalisation of
$D_{xx} \sim 10^{28} \text{cm}^{2} \, \text{s}^{-1}$. The value of
$\delta \sim 1/3$ corresponding to Kolmogorov turbulence is consistent
with observations of nuclear secondary-to-primary ratios only in
models with reacceleration, otherwise $\delta \sim 0.5$ fits
better~\cite{Trotta:2010mx} (see however \cite{Maurin:2010zp} who
report that an even larger value $\delta \sim 0.8$ is
possible). Therefore, for the energies relevant here, i.e. $E \lesssim
100 \, \text{GeV}$, any structure in the distribution of CR electron
sources is washed out by diffusion, such that the CR electron
distribution is quite smooth on scales of $\lesssim 100 \, \text{pc}$.

On large ($\gtrsim 1 \, \text{kpc}$) scales, however, there are
variations in both the CR electron distribution and the GMF. The
variation of the electron density is determined by the spatial
distribution of CR electron sources and the details of the propagation
model, e.g. shape and extent of the propagation volume, diffusion
coefficients and speed(s) of convective wind(s) if any. For the GMF,
the ordered component is expected to follow the Galactic spiral arms
(as is observed in other galaxies e.g. M82, M51, M81, NGC1068 and
NGC6946), while the RMS value of its turbulent component is expected
to decrease with distance, both from the Galactic centre and from the
disk~\cite{Beck:2013bxa}.

Attempts at modelling the diffuse synchrotron background have focussed
on the distribution of emissivity on either very small or very large
scales. It is
known~\cite{Chepurnov:1998zz,Cho:2002qk,Cho:2010kw,Lazarian:2011ik}
that (assuming a statistically isotropic sky) the power law of the ISM
turbulence is reflected in a broken power-law for the APS: below some
critical multipole $l_\text{cr}$, $\mathcal{C}_l \propto l^{-1}$,
whereas for larger multipoles (smaller angular scales), $\mathcal{C}_l
\propto l^{-\alpha}$. The critical multipole depends on the `outer
scale' of turbulence $L$ and the scale height $R$ of the turbulent
medium: $l_\text{cr} \sim 2 \pi R/L$. However, comparison with
observations is hampered by the fact that the observed synchrotron sky
is \emph{not} statistically isotropic, due to large-scale variations
of emissivity. Studies of MHD turbulence in the APS have therefore
constrained themselves to patches of the sky, e.g. by cutting out
bands around the Galactic plane. Clearly, this cannot describe the APS
down to small $l$; in fact, attempts to fit the APS by a broken power
law are able to reproduce the observed behaviour for $l \lesssim 20$,
$l (l+1) \mathcal{C}_l \sim \text{constant}$, only by assuming a small
scale height, $R \lesssim 1 \, \text{kpc}$~\cite{Regis:2011ji}, which
is quite
unrealistic~\cite{Strong:2011wd,Bringmann:2011py,DiBernardo:2012zu}.

On large scales, two approaches have been adopted in the literature:
Motivated by the need for a realistic model of the polarised emission,
one set of
studies~\cite{Sun:2007mx,Sun:2010sm,Fauvet:2010xq,Page:2006hz} employ
sophisticated models (of the ordered component) of the GMF constrained
by detailed analyses of rotation measures (RM) and starlight
polarisation, but make rather simplistic assumptions for the spatial
distribution of relativistic electrons. The other set of
studies~\cite{Orlando:2009xg,Jaffe:2011qw,Bringmann:2011py,Jaffe:2013yi}
start with a detailed model for CR propagation, reproducing locally
measured spectra of protons, nuclei and electrons, and fix the (random
component) of the GMF by fitting to the strength of the synchrotron
radiation. Most of the emission on large scales would appear to be
produced by the random component --- in fact the magnitude of the
ordered component as obtained from RM data is too weak to explain the
radio flux, as has been recognised for some years
\cite{Badhwar:1977sv}. However, the random component has \emph{not}
been modelled as a true random field but only in terms of its RMS
value.  While this modelling of the large-scale component can
reproduce the global behaviour, there is mismatch on intermediate
scales of tens of degrees. To downplay this, angular profiles of the
radio background are usually averaged over large parts of the sky
before comparison with data. The remaining discrepancies on smaller
scales are usually attributed to ``local structures'' but there have
been few attempts to model these.

\begin{figure}[ht]
\begin{minipage}[t]{0.5\linewidth}
\centering
\includegraphics[width=\textwidth]{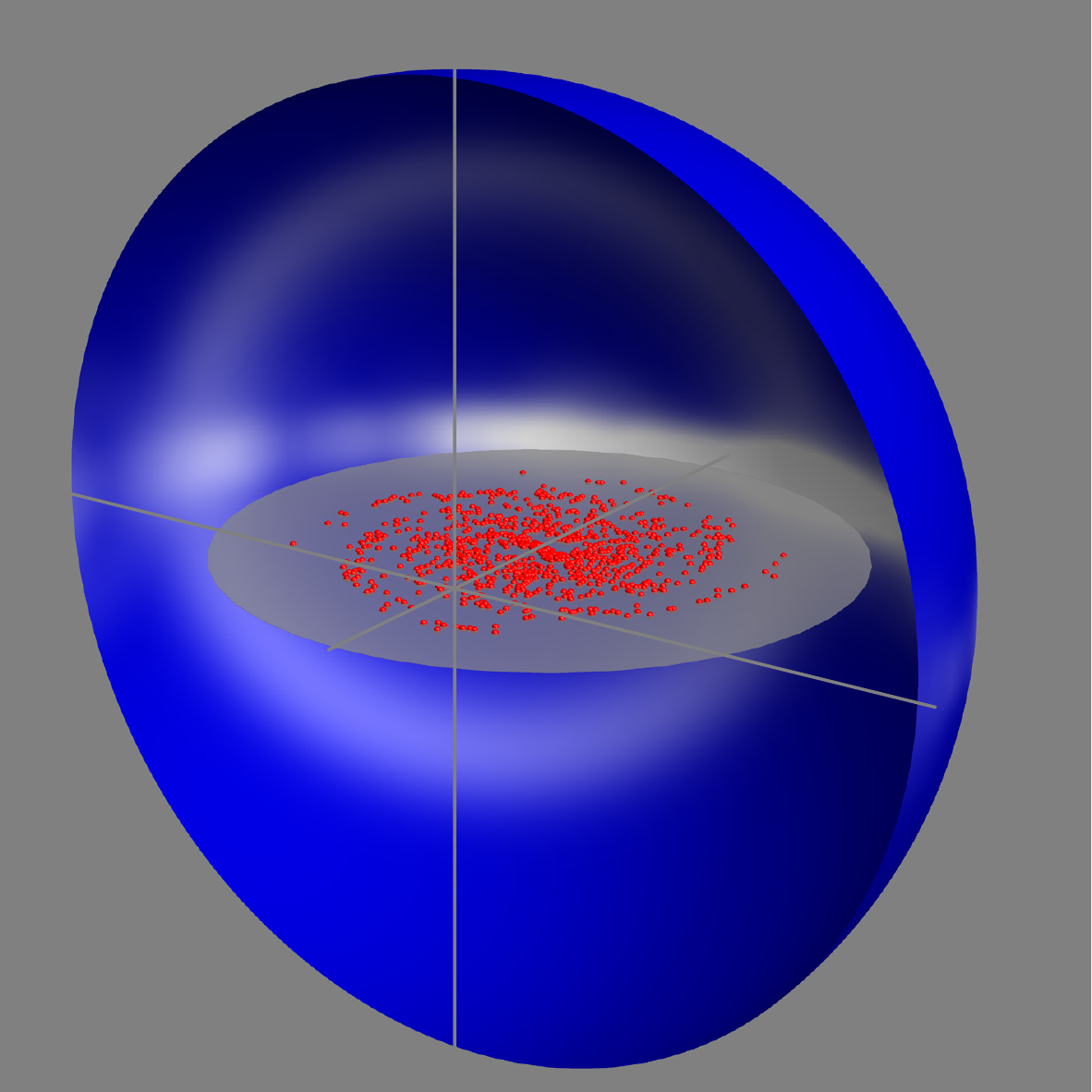}
\caption{Sketch of the Galactic disk with shells of old supernova
  remnants and their skymap, i.e. their projection onto the celestial
  (hemi)sphere, in order to indicate the origin of `loop' features in
  the radio sky. The (helio-centric) Galactic coordinate system is
  shown by the grey lines.}
\label{fig:figure1}
\end{minipage}
\hspace{0.5cm}
\begin{minipage}[t]{0.5\linewidth}
\centering
\includegraphics[width=\textwidth]{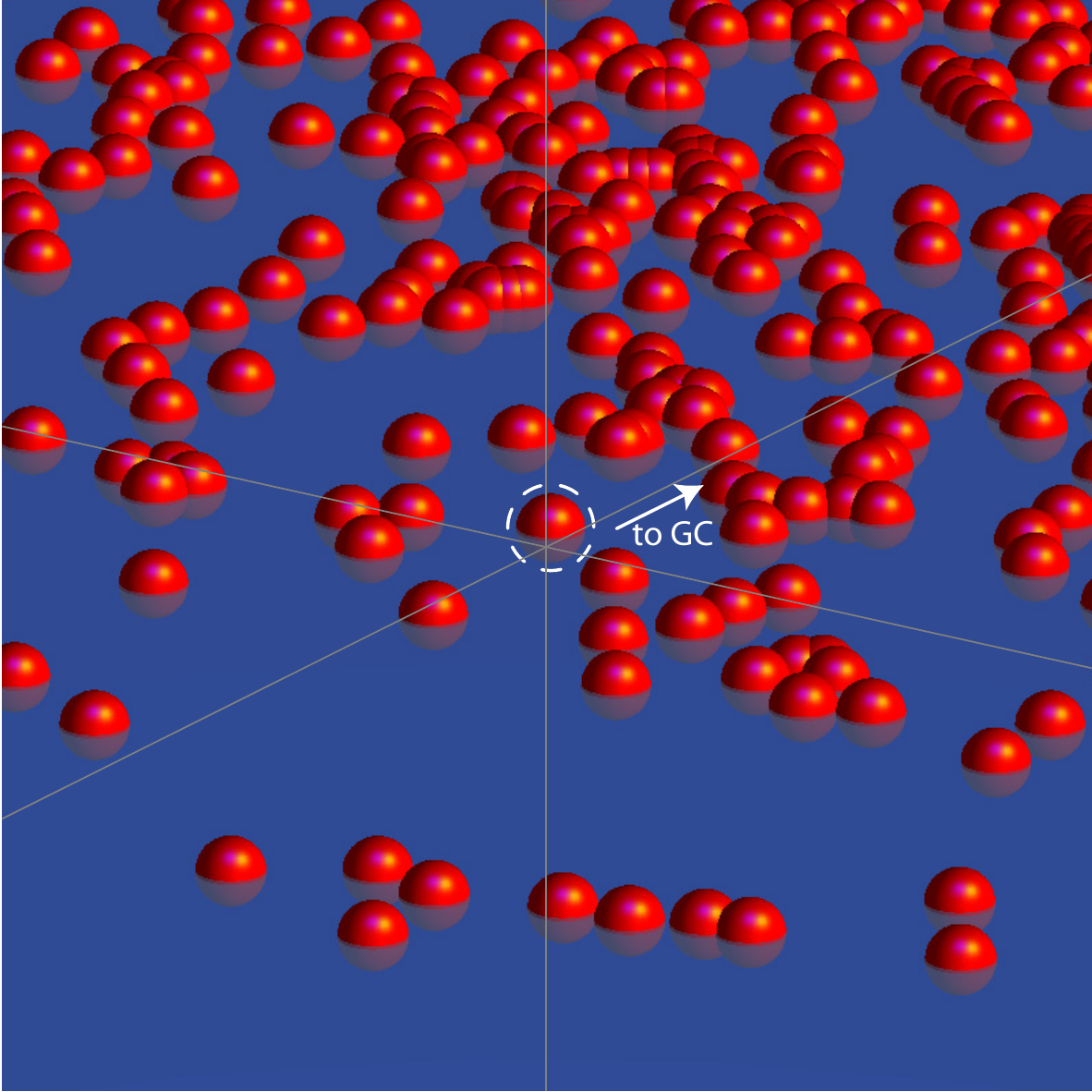}
\caption{Zoom into the inner few kiloparsecs of the Galactic disk
  around the position of the Sun. The nearest old SNR which is
  responsible for the largest `loop' feature on the radio sky is
  circled.}
\label{fig:figure2}
\end{minipage}
\end{figure}

It is thus necessary to model the emission consistently on both large
and small angular scales. There must also be additional contributions
at intermediate angular scales from the $\sim 1000$ old supernova
remnants (SNRs) in the Galaxy which have expanded to large sizes of
$\mathcal{O}(100) \, \text{pc}$ in the hot tenuous ISM before cooling
and turning radiative \cite{1977ApJ...218..148M}. Although these old
SNRs may not actively accelerate CRs, they do compress the GMF in
their `supershells' by large factors, and with it the tied CR
electrons, which would moreover be then betatron accelerated to higher
energies \cite{1962MNRAS.124..125V}. As the synchrotron emissivity
scales as the square of the magnetic field strength and of the
\emph{correlated} increase in CR electron energy
\cite{1974ApJ...189...51C}, the radio emission from such old SNR
shells can be significantly amplified. It has been argued that if most
of the Galactic `diffuse' background arises from such (unresolved) old
SNRs then the absolute emissivity can be explained with an average
magnetic field strength of $\sim 2\,\mu$G, consistent with the RM
measurements \cite{Sarkar:1982zz}. These large old SNRs may themselves
generate the observed fluctuating component of the GMF. Given that
they pervade the whole Galaxy and can be found at distances between
hundreds of pc to tens of kpc, we expect them to contribute at angular
scales between a few degrees up to the full sky (see
Figs.~\ref{fig:figure1} and~\ref{fig:figure2}).

In this paper, we attempt to model the radio sky on a wide range of
scales, taking into account the large-scale contribution of CR
electrons and of the GMF, the variation of the GMF due to the
turbulence cascade in the ISM and the effect of the presence of large
radiative shells of old SNRs. The observable we model is not the
all-sky radio map itself but its APS, i.e. the distribution of
$\mathcal{C}_l$ with $l$. Modelling the APS instead of sky maps has
several advantages; firstly, the cascading of energy from the outer
scale $L$ to the dissipation scale as well as the spatial distribution
of old SNRs is a stochastic process of which the observed Galaxy
represents only one particular realisation. What can be predicted are
the ensemble averages which may to a certain degree differ from the
observed Galaxy. The APS is a representation of this stochastic
process that reduces the variation between different realisations
by efficiently averaging over many lines of sight. Secondly, the APS
presents a representation of the information sorted by angular size
and also encodes the symmetry properties of the sky map. Finally, for
CMB studies (and also a particular class of indirect searches for DM
annihilation signals), the observable is also the APS and it is
therefore favourable to perform the foreground subtraction not on sky
maps but \emph{directly} on the APS.

The remainder of this paper is organised as follows. In
\S~\ref{sec:Haslam} we discuss the APS of the `Haslam' 408 MHz all-sky
survey~\cite{Haslam:1982aa}. In \S~\ref{sec:Modelling} we present our
model for the synchrotron background, discussing in detail each
component (large-scale smooth emission, small-scale turbulence,
intermediate scale old SNR shells) in turn. We demonstrate a good
match of our model to the observed APS. In subsequent papers we will
present the match to the observed frequency spectrum for different
regions of the sky, as well as to the RM sky maps.

\section{The observed angular power spectrum}
\label{sec:Haslam}

We begin by computing the APS of the
\HaslamASS{}~\cite{Haslam:1982aa}. Throughout this study, we used a
verison of the map that has been corrected for scanning artefacts and
with strong point sources
removed.\footnote{\url{http://lambda.gsfc.nasa.gov/product/foreground/fg_haslam_get.cfm}}
It is possible that unresolved point sources still contribute (in
particular at high multipoles), however we will argue later that this
is most likely negligible. Furthermore, as this is the only full-sky
point-source subtracted map that is publicly available, we restrict
ourselves, for the time being, to this one frequency. Important
additional information is encoded in the frequency-dependence of the
sky maps (or equivalently the APS), because of the different (and
spatially varying) frequency dependence of the various physical
components contributing to the diffuse emission, see
\S~\ref{sec:Modelling}. This is particularly true for synchrotron
emission by relativistic electrons, given that the spatial
distribution of the radiating CR electrons varies with energy due to
their energy-dependent diffusion and energy loss rates. Extrapolating
the APS (or sky maps, for that matter) between widely separated
frequencies is therefore quite
non-trivial~\cite{Mertsch:2010ga}. However, if successful in
developing a convincing physical model for the APS, we can be of
course more confident in predicting the APS at other frequencies. It
would therefore be desirable to test and (if necessary) calibrate our
model by comparing to the APS at different frequencies. Here we adopt
the complementary approach of fixing all the parameters using other
observations, which allows for testing and dissecting the underlying
(model) assumptions.

\begin{figure}[ht]
\centering
\includegraphics[scale=1]{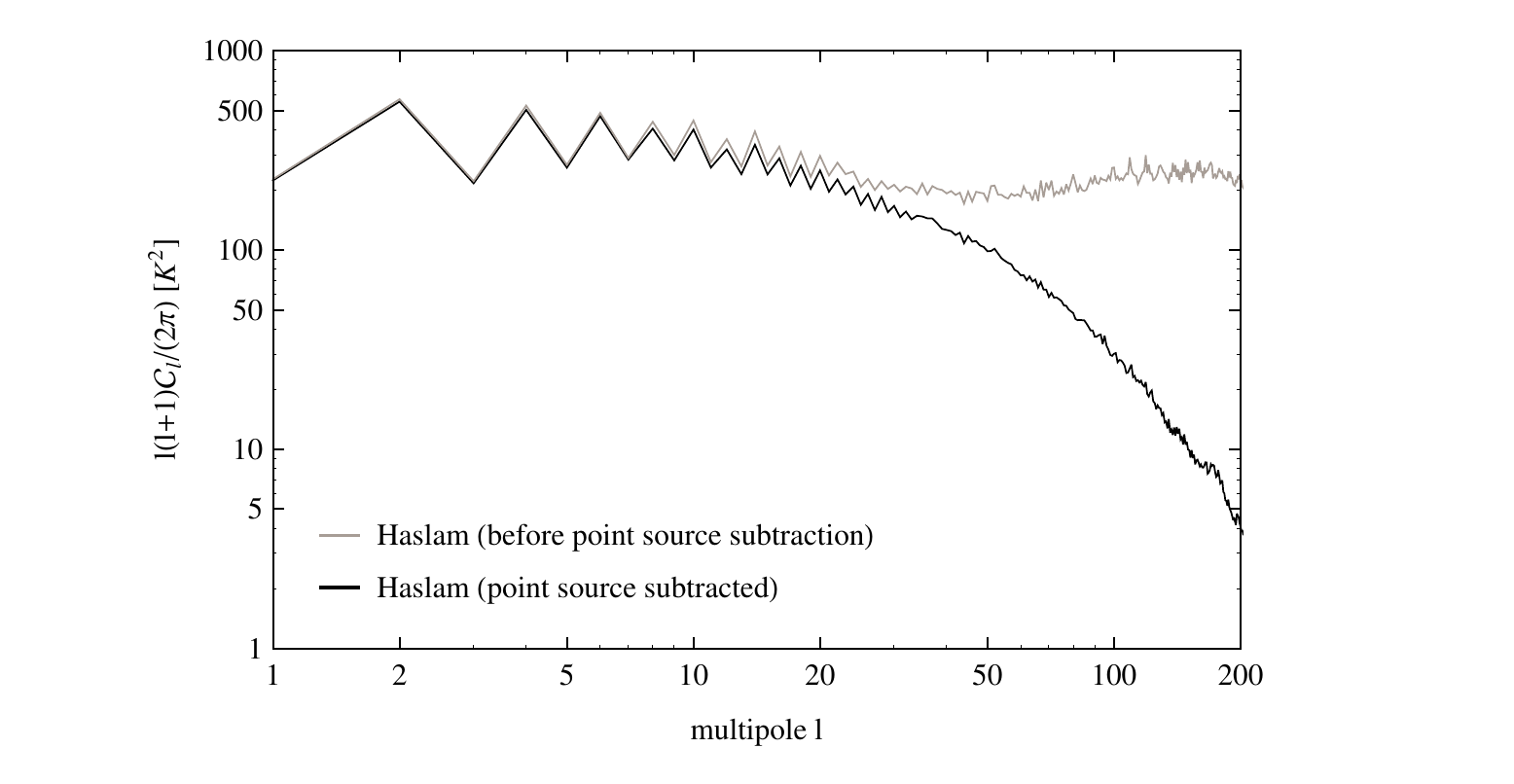}
\caption{The APS of the \HaslamASS{}, both before and
  after correction for scanning artefacts and subtraction
  of point sources.}
\label{fig:Haslam}
\end{figure}

In Fig.~\ref{fig:Haslam}, we show the APS of the
\HaslamASS{}~calculated using \texttt{anafast} from the
\texttt{HEALPix}-suite~\cite{Gorski:2004by}. For cosmological studies,
it is customary to plot the $\mathcal{C}_l$ multiplied by $l(l+1)/(2
\pi)$ in order to facilitate comparison with a Harrison-Zeldovich
primordial spectrum for which $\mathcal{C}_l \propto 1/l(l+1)$. Doing
so is still helpful in the present context since multiplying
$\mathcal{C}_l$ by $l(l+1) \sim l(2l+1)$ directly displays the power
contained in logarithmic intervals of $l$.

It is seen that most of the power in the \Haslam{} map is present at
multipoles $l \lesssim 10$, corresponding to angular scales $\gtrsim
20^\circ$. The most prominent feature in the APS, however, is the
disparity between even and odd modes --- a consequence of the
(approximate) symmetry properties of the synchrotron map, in
particular the symmetry between the northern and southern (Galactic)
sky and the invariance with respect to shifts in longitude: If a sky
map $J(\theta, \phi)$ is perfectly symmetric with respect to the
Galactic plane, $J(\theta, \phi) \equiv J (\pi-\theta, \phi)$, then
since the spherical harmonics satisfy $Y_{lm}(\pi-\theta, \phi) =
(-1)^{l+m} Y_{lm}(\theta, \phi)$, it follows that $a_{lm} = (-1)^{l+m}
a_{lm}$. If, in addition, we assume that the synchrotron flux is
independent of longitude $\phi$, then all $a_{lm}$ vanish except when
$m=0$. This results in the $a_{lm}$ for even $l$ (and $m=0$) being the
only non-vanishing components, and therefore also the $\mathcal{C}_l$s
with even $l$. However, as the above assumptions are satisfied only
approximately (in particular, invariance with respect to shifts in
longitude), the symmetry is partially broken and the $\mathcal{C}_l$s
with odd $l$ become populated. Thus the ratio of odd to even
$\mathcal{C}_l$s is a measure of how much the Galactic plane symmetry
is broken, and by how much the flux varies with longitude.

It is also apparent from Fig.~\ref{fig:Haslam} that this odd-even
disparity does not extend to arbitrarily high multipoles. In fact,
between $l=30$ and $40$, the saw-tooth structure largely disappears,
pointing at a different component dominating on smaller
scales. Moreover the APS at large ($l \gtrsim 50$) multipoles becomes
rather smooth; this might seem surprising given how apparently
stochastic structure is present in the sky maps on small
scales. However, as mentioned already, the APS is formed at large $l$
by averaging over many different $m$-modes and correspondingly many
different lines-of-sight (LoS), which leads to an efficient averaging
of the stochastic information (i.e. the distribution of turbulent
eddies or shells of old SNRs) and carves out the physically meaningful
information that can actually be predicted, viz. their statistical
expectation values.

\section{Modelling of the angular power spectrum}
\label{sec:Modelling}

\subsection{Definitions}
\label{sec:definitions}

\begin{figure}[tbh]
\begin{tabular}{ccc}
\includegraphics[width=0.3\textwidth]{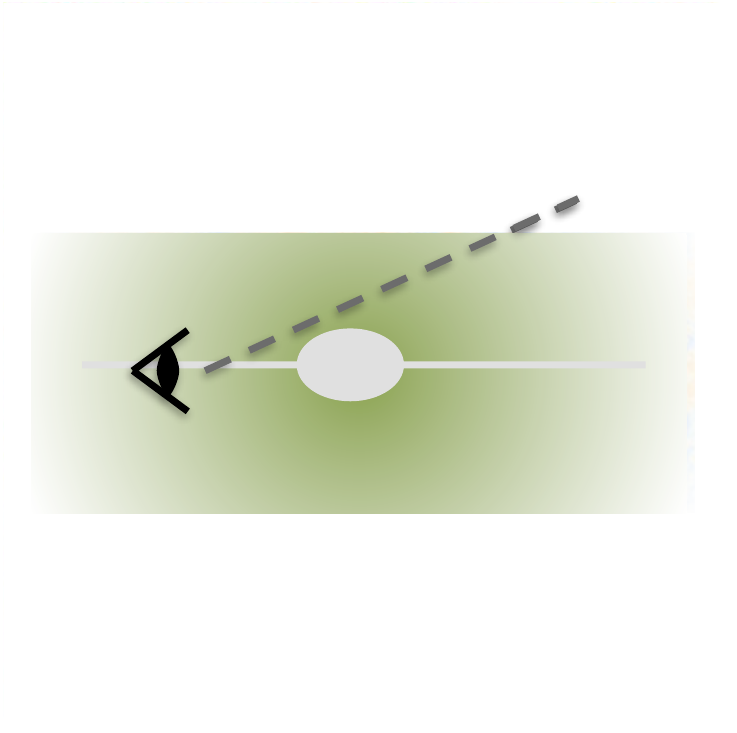} &
\includegraphics[width=0.3\textwidth]{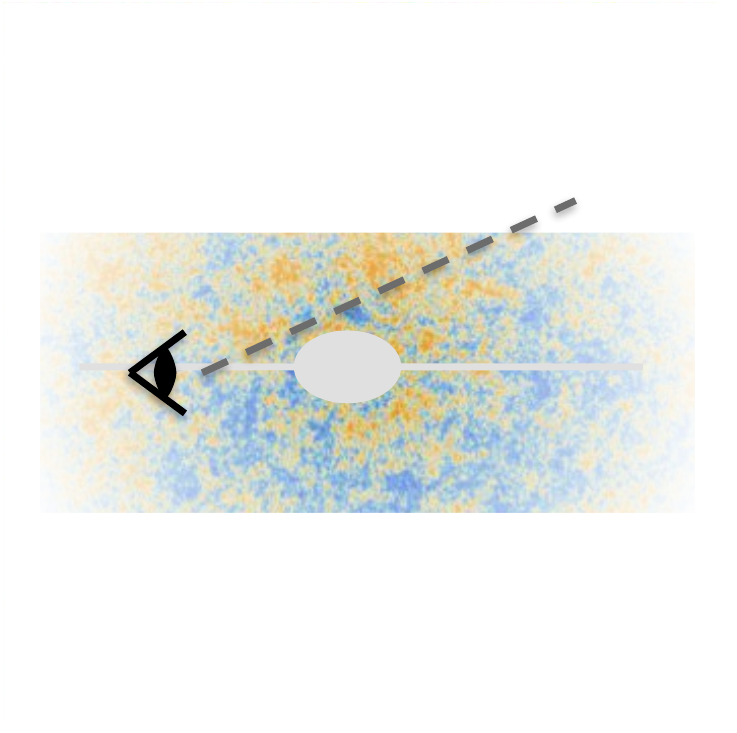} &
\includegraphics[width=0.3\textwidth]{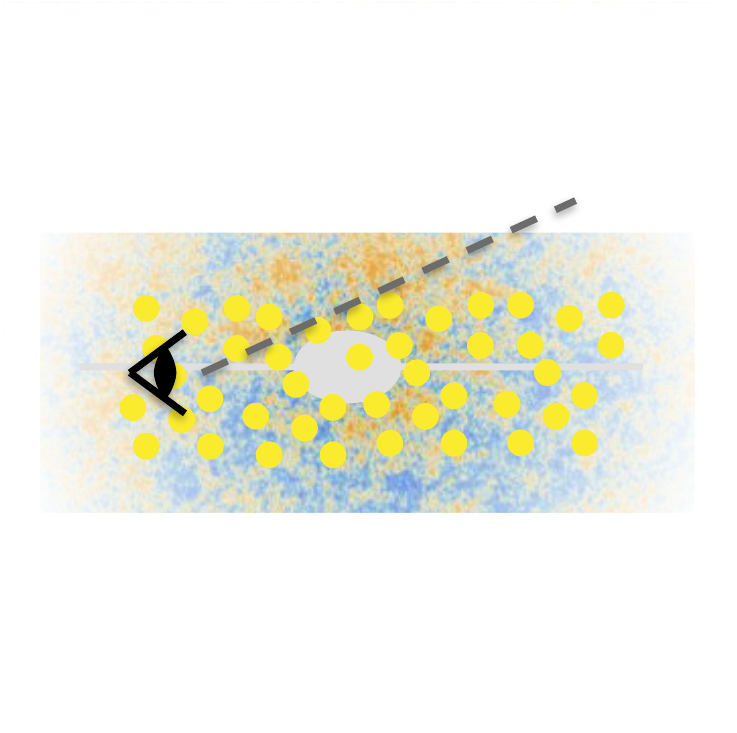} \\
(a) & (b) & (c)
\end{tabular}
\caption{Different setups for calculations: (a) Smooth, large-scale
  emissivity with an off-center observer as considered in the
  \texttt{GALPROP} code (\S~\ref{sec:large-scale}); (b) Turbulent halo
  dressed with the large-scale emissivity of the \texttt{GALPROP} code
  as in our MC calculation (\S~\ref{sec:small-scale}); (c) As in panel (b)
  but with an additional contribution arising from a large number of
  old SNR shells (\S~\ref{sec:shells}).}
\label{fig:setups}
\end{figure}

The power emitted at frequency $\nu$ per unit solid angle and
frequency by an electron of energy $E$ in a magnetic field $\vec{B}$
is~\cite{Longair:2010aa}:
\begin{equation}
\varepsilon(\nu, E, \vec{B}) = \frac{\sqrt{3} e^3 B_{\perp}}{8 \pi^2
  \varepsilon_0 c m_e} \left( \frac{\nu}{\nu_c} \right)
\int_{\nu/\nu_c}^{\infty} \dd x \, K_{5/3} (x) \quad
\text{with} \quad \nu_c = \frac{3}{2} \gamma^2 \frac{e B_{\perp}}{2
  \pi m_e} \, ,
\label{eqn:MonoenergeticPower}
\end{equation}
where $B_{\perp}$ is the component of $\vec{B}$ perpendicular to the
LoS, $e$, $m_e$ and $\gamma$ are, respectively, the electron charge,
mass and relativistic gamma factor and $K_{5/3}$ is the Bessel
function of degree $5/3$. Given a number $n_e(\vec{r}, E)$ of
electrons per unit energy and volume at position $\vec{r}$ and with
energy $E$, the total synchrotron power
$\hat{\varepsilon}(\nu,\vec{r})$ is then
\begin{equation}
\hat{\varepsilon}(\nu, \vec{r}) = \int \dd E \, n_e(E, \vec{r})
\varepsilon(\nu, E, \vec{B}(\vec{r})).
\label{eqn:SpectrumPower}
\end{equation}
For a power law electron spectrum $n_e(E, \vec{r}) \propto
E^{-\gamma}$, $\hat{\varepsilon}(\nu, \vec{r}) \propto
\nu^{-(\gamma-1)/2} B_{\perp}^{(\gamma+1)/2} (\vec{r})$. For the
locally measured $\gamma \simeq 3$, this gives:
$\hat{\varepsilon}(\nu, \vec{r}) \propto \nu^{-1} B_{\perp}^{2}
(\vec{r})$.

The flux $J(\nu;\theta,\phi)$ observed in a particular direction
$(\theta,\phi)$ on the sky is obtained by integrating the synchrotron
emissivity $\hat{\varepsilon}(\nu,\vec{r})$ over the corresponding
LoS, $\vec{r}(s,\theta,\phi)$,
\begin{equation}
J(\nu;\theta,\phi) = \int_0^{\infty} \dd s \, \hat{\varepsilon} (\nu,
\vec{r}(s,\theta,\phi)) = \int_0^{\infty} \dd s \int \dd E \, n_e
\big( E, \vec{r}(s,\theta,\phi) \big) \varepsilon \big( \nu, E,
\vec{B}(\vec{r}(s,\theta,\phi)) \big) \, .
\label{eqn:LOSintegration}
\end{equation}

In the following, we model the APS of the diffuse Galactic radio
background, taking into account dynamics over a wide physical range:
the emissivity of the ISM with variations on large (see
Fig~\ref{fig:setups}a) and small (see Fig~\ref{fig:setups}b) scales as
well as the shells of old supernova remnants (see Fig~\ref{fig:setups}c).
We also take into account free-free emission as well as the
presence of unsubtracted point sources, although this is not the main
thrust of this work.

\subsection{Large-scale variation of synchrotron emissivity}
\label{sec:large-scale}

We start by modelling the GMF on scales much larger than $L$ of ${\cal
  O}(100) \, \text{pc}$, the assumed `outer scale' of turbulence. For
this we use the \texttt{GALPROP} code~\cite{Vladimirov:2010aq} which
numerically solves the transport equation for Galactic cosmic rays
(GCRs), taking into account the distribution of the sources (as
determined from e.g.  SNR or pulsar surveys), the spatial diffusion of
CRs in the turbulent GMF, their convection by Galactic winds, various
energy loss processes and `reacceleration' due to diffusion in
momentum space. We adopt the ``pure diffusion model''
\texttt{z04LMPD}~\cite{Strong:2010pr}, which is known to reproduce
both the locally measured electron spectrum as well as the synchrotron
spectrum averaged over large parts of the sky out of the Galactic
plane. This model's parameter values can also reproduce locally
measured GCR fluxes and secondary-to-primary ratios. While the height
of the CR halo is important for the large-scale distribution of the
synchrotron emitting electrons and can therefore affect the lower
multipoles, we have chosen to fix it to the best-fit-value
$z_{\text{max}} = 4 \, \text{kpc}$~\cite{Trotta:2010mx} as we are not carrying out a full
parameter study.  Although the choice $\delta = 0.5$ in
\texttt{z04LMPD} is somewhat at odds with our assumed Kolmogorv
spectrum for the ISM turbulence (which predicts $\delta = 1/3$), we
chose to retain it as a change of $\delta$ can easily be absorbed into
the source spectral indices without changing the primary spectra too
much. Note that as the assumed source distribution varies over
distances $\gtrsim L$, so will the electron density. Rather than
describe the turbulent GMF fully, we consider only its RMS value,
averaged over ensembles of turbulent GMFs. For the normalisation and
spatial dependence we adopt the model that is
stated~\cite{Strong:2011wd} to give a good fit to global angular
profiles of the \Haslam~all-sky survey, viz.
\begin{equation}
B_{\text{rms}}(r,z) = B_0 \, \ee^{-r/\rho - |z|/\xi} \, ,
\label{eqn:TurbFieldPosDep}
\end{equation}
with $B_0 = 7.5 \, \mu\text{G}$, $\rho=30 \, \text{kpc}$ and $\xi=4 \,
\text{kpc}$. Later we increase $\rho$ to $100 \, \text{kpc}$
(i.e. essentially no variation with radius) to better reproduce the
symmetry properties of the APS. We chose to ignore the ordered
component altogether for the time being as its magnitude is much
smaller than the RMS of the turbulent component. Note in this context, 
that due to the physically unavoidable correlations between magnetic 
field strength and electron density, the regular field as determined from 
synchrotron observations alone is likely 
overestimated~\cite{Beck:2003dd}, as was first pointed out many years 
ago~\cite{1974ApJ...189...51C,Sarkar:1982zz}.

It has been recognised for some time \cite{Longair:2010aa} that the
spectrum of the diffuse radio background softens between at few
hundreds of MHz and a few GHz, which probably reflects a break in the
power-law describing the underlying electron spectrum. We
follow~\cite{Jaffe:2009hh,Strong:2011wd,DiBernardo:2012zu} in adopting
a spectral break in the (source) electron spectrum at $\mathcal{R}_1 =
4 \, \text{GV}$ (which has been also found to be necessary for fitting
the spectra of protons and
nuclei~\cite{Trotta:2010mx,Kachelriess:2012fz}). This break has also
been attributed to a change in the energy dependence of the diffusion
coefficient \cite{Ptuskin:2005ax,Blasi:2012yr} but as the effect on
the propagated spectrum is similar, we can remain agnostic as to its
origin. The second break at $\mathcal{R}_2 = 50 \, \text{GV}$ is
probably not due to either cause but possibly the effect of an
additional, harder contribution to the electron flux. (If the sources
also accelerate positrons with a similar spectrum, this would explain
the observed rise in the positron fraction~\cite{Ahlers:2009ae}.) The
parameters adopted are shown in Table~\ref{tbl:GALPROP_parameters}.

The synchrotron sky map is readily obtained from the \texttt{GALPROP}
code in the \texttt{HEALPix} scheme and its APS is again calculated
using the \texttt{HEALPix} suite.

\begin{table}[tbh]
\centering
\begin{tabular}{| l | l | l |}
\hline Description & Parameter & Value \\ \hline \hline Spatial
diffusion coefficient at 4~GV & $D_0$ & $3.4 \times 10^{28} \,
\text{cm}^{2} \text{s}^{-1}$ \\ Power lax index of the spatial
diffusion coefficient & $\delta$ & 0.5 \\ Half-height of CR
propagation volume & $z_\text{max}$ & 4~kpc \\
\hline \hline
\multirow{2}{*}{Source distribution} & \multicolumn{2}{l|}{\textit{pulsar-like}~\cite{Lorimer:2003qc}, but held}  \\ 				
& \multicolumn{2}{l|}{constant for $r>10 \, \text{kpc}$}  \\													 \hline	
\multirow{3}{12em}{Source spectral indices} for $\mathcal{R} \leq \mathcal{R}_1$								& $\Gamma_1$ 	 	& $1.6$ \\
\multirow{1}{12em}{} for $\mathcal{R}_1 < \mathcal{R} \leq \mathcal{R}_2$					 											& $\Gamma_2$ 	 	& $2.5$ \\
\multirow{1}{12em}{} for $\mathcal{R}_2 < \mathcal{R}$					 											& $\Gamma_3$ 	 	& $2.2$ \\
\multirow{2}{9.3em}{Break rigidities} & $\mathcal{R}_1$ & $4 \, \text{GV}$ \\
& $\mathcal{R}_2$ & $50 \, \text{GV}$ \\
\hline \hline
Magnetic field strength at the Galactic centre & $B_0$ & $7.5 \, \mu\text{G}$ \\
Scale height in radial direction	& $\rho$ & 30~kpc, 100~kpc \\
Scale height in direction perpendicular to Galactic plane	& $\xi$ & 4~kpc \\
\hline
\end{tabular}
\caption{Parameters of the propagation and magnetic field
  model~\cite{Strong:2010pr,Strong:2011wd} adopted for the computation
  of the large-scale component of the synchrotron emission.}
\label{tbl:GALPROP_parameters}
\end{table}

\subsection{Small-scale variation of synchrotron emissivity}
\label{sec:small-scale}

Clearly the assumption that the GMF has structure only on scales
larger than $L$ of ${\cal O}(100) \, \text{pc}$ is
unrealistic. Firstly, the magnitude of the GMF as determined from
Faraday rotation (which depends on the large-scale component, i.e. the
GMF averaged over LoS $\gtrsim L$) is too small \cite{Badhwar:1977sv}
to reproduce the overall synchrotron flux (which depends on the mean
square average value along the LoS). Hence, the variation on scales
smaller than $L$ cannot be neglected. Secondly, a turbulent GMF field
is also inferred from the fact that the observed anisotropy in the
arrival directions of GCRs is very small: around $10^{-3}$ up to $100
\, \text{TeV}$ and around $10^{-2}$ at $10 \, \text{PeV}$
\cite{Ptuskin:2012vs}. It is therefore believed that the dominant
process of CR transport in the Galaxy is diffusion due to resonant
wave-particle interactions --- this requires a turbulent magnetic
field on the scale of the particle gyroradius, e.g. $\sim 10 \,
\text{pc}$ for $E = 10 \, \text{PeV}$ in a $\mu \text{G}$ field.

Under some simplifying assumptions the APS of the synchrotron emission
from a turbulent magnetic field can be computed
analytically~\cite{Chepurnov:1998zz}. Let the magnetic field be a
Gaussian random field with an ensemble average that is only a function
of the distance $s$ from the observer: $\langle \vec{B}^2 \rangle
\propto w(s)$ --- we call this the statistically isotropic setup (see
also Fig.~\ref{fig:StatIsoSetup}). Let us approximate the electron
spectrum as a power law $n(E) \propto E^{-3}$ which is close to the
locally observed electron spectrum. The emissivity is then
proportional to the square of the magnetic field perpendicular to the
LoS which on average --- assuming statistical isotropy --- is 2/3 of
the square of the total magnetic field. It turns out that the 3D power
spectrum of the random field (i.e. the Fourier transform of the
two-point correlation function) is then reflected in the APS of the
resulting synchrotron sky map~\cite{Chepurnov:1998zz}. In fact, if we
assume a Kolmogorov power law with a cut-off at small wavenumber $k_0$
(corresponding to an outer scale of turbulence: $L \sim 1/k_0$),
\begin{equation}
\frac{\dd B^2}{\dd^3 k} = \mathcal{F}_0^2 k^{-11/3} \ee^{- k_0^2/k^2} \, .
\label{eqn:DefdF2d3k}
\end{equation}
If we further adopt $w(s) = \ee^{-s^2/R^2}$, then for $l \gg 1$, the
APS is a broken power law in $l$:
\begin{equation}
\mathcal{C}_l = 2.96 \times 10^3 \frac{\mathcal{F}_0^4
  R^{8/3}}{k_0^{2/3}} \frac{1+3.21 \ee^{-1.43 \, l/(k_0 R)}}{l^{11/3}
  \left(1 + 5.0 \left( \frac{k_0 R}{l} \right)^2 \right)^{4/3}} \,
, \label{eqn:APSsmallscale}
\end{equation}
with power law indices $-1$ ($-11/3$) for $l \lesssim l_{\text{cr}}$
($l \gtrsim l_{\text{cr}}$). The break occurs around a critical
multipole that depends on the path length $R$ and the outer scale of
turbulence $L$ as $l_\text{cr} \approx (2\pi
R/L)$~\cite{Chepurnov:1998zz,Cho:2002qk,Cho:2010kw}.

It has been argued that the \emph{observed} APS follows a power law at
large $l$ with spectral index $-11/3$ due to this small-scale
variation of synchrotron
emissivity~\cite{Chepurnov:1998zz,Cho:2002qk,Cho:2010kw,Regis:2011ji}. However,
for this to hold the normalisation of the observed high-$l$ component
should also be reproduced. In fact, in the statistically isotropic
setup, there is a close relation between the monopole $\mathcal{C}_0$
and the power of the higher ($l>0$) multipoles. It is shown in
Appendix~\ref{sec:Monopole} however that the power in the monopole is
always many orders of magnitude larger than the power contributed by
the rest of the APS. Therefore, fixing the normalisation
$\mathcal{F}^2_0$ in eq.~(\ref{eqn:DefdF2d3k}) to reproduce the APS at
large $l$ from eq.~(\ref{eqn:APSsmallscale}) will imply too large a
monopole. We conclude that the magnetic turbulence can contribute very
little to the APS at large $l$.

Of course the real question is how the picture changes when we
consider a more realistic setup, viz. an off-center observer in a
medium with small-scale turbulence, that is dressed with the overall
large-scale emissivity of the \texttt{GALPROP} run from
\S~\ref{sec:large-scale} (cf. Fig.~\ref{fig:setups}b). One would then
expect the APS to be more-or-less the sum of the large-scale and the
small-scale contributions as these are on disjoint scales. For the
small-scale component, one would need to average the APS over a
distribution of scale-lengths $R$ since the scale length in the
off-center setup depends on the direction~\cite{Cho:2002qk}. Moreover
since the large-scale emissivity is no longer isotropic, the monopole
computed in (\ref{eqn:monopole}) gets spread out into the lowest
multipoles and we would therefore expect the large-scale power in
these lowest multipoles to dominate over the small-scale power at the
higher multipoles.

\begin{figure}[ht]
\centering
\includegraphics[scale=1]{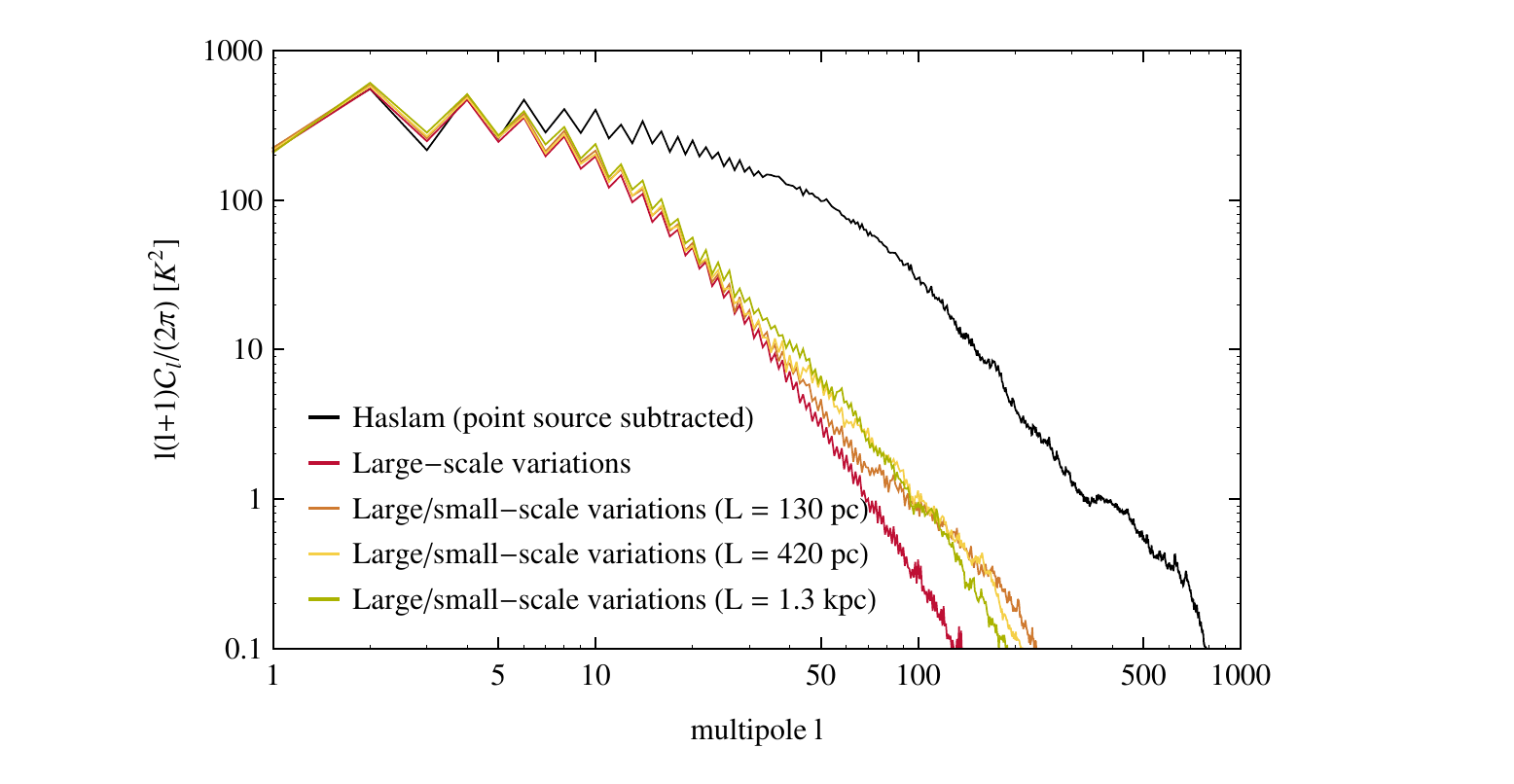}
\caption{The APS of radio emission due to small-scale turbulence in
  the Galactic magnetic field for different choices of the `outer
  scale' $L$.}
\label{fig:APS_Turb}
\end{figure}

We have computed the sky map and its APS numerically by a Monte Carlo
calculation. To that end we have prepared a turbulent magnetic field
by superposing hundreds of plane waves evenly spaced
logarithmically in $k$ with amplitudes chosen so as to reproduce a
power law power spectrum $\propto k^{-11/3}$ with unit variance for
$k>2 \pi/L$~\cite{Giacalone:1999aa}. In particular, 
we adopt a coherence length $L = 420 \,\text{pc}$. This
statistically isotropic Gaussian random field $\vec{b}(\vec{r})$ is
then considered as the deviation of the turbulent magnetic field
$\vec{B}(\vec{r})$ from its ensemble average $B_{\text{rms}} (r,z)$
(eq.~\ref{eqn:TurbFieldPosDep}), i.e. $\vec{B}(\vec{r}) =
B_{\text{rms}} (r,z) \vec{b}(\vec{r})$. Given the (smooth) emissivity
(cf. \S~\ref{sec:small-scale}) calculated by \texttt{GALPROP},
$\hat{\varepsilon}(\nu;B_{\text{rms}}(r,z))$, we can now compute the
emissivity from the fully turbulent field by using the scaling of the
emissivity and critical frequency with the magnetic field $B$
(cf. eq.~\ref{eqn:MonoenergeticPower}),
\begin{equation}
\varepsilon(\nu; \beta B) = \beta \varepsilon(\nu/\beta; B) \, ,
\label{eqn:enhancement}
\end{equation}
where $\beta = |\vec{b}_{\perp}(\vec{r})|$. We generate a sky map by
integrating this emissivity along LoS's over the whole sky and
calculate its APS as above. In Fig.~\ref{fig:APS_Turb} we show that
the dependence on the value of $L$ is rather mild.

\subsection{Free-free emission}
\label{sec:freefree}

Free-free emission is the bremsstrahlung of thermal electrons in the
ISM. For an electron temperature $T_e$ (in units of degrees Kelvin),
the optical depth is given by~\cite{Dickinson:2003vp}
\begin{equation}
\tau_c = 3.014 \times 10^{-2} T_e^{-3/2} \nu^{-2} \left( \ln \left[
  4.955 \times 10^{-2} \nu^{-1} \right] + 1.5 \ln T_e \right)
\int_{\text{LoS}} \dd \vec{r} \, n_e n_i
\end{equation}
and the brightness temperature is $\propto \tau_c T_e$. A reasonable
approximation is
\begin{equation}
\tau_c \propto T_e^{-1.35} \nu^{-2.1} \int_{\text{LoS}} \dd \vec{r} \,
n_e n_i \, ,
\end{equation}
and the brightness temperature then factorises into a frequency part
($\propto \nu^{-2.1}$) and a spatial/angular part.

Consequently, unlike synchrotron emission the spatial distribution of
thermal bremsstrahlung emissivity is not expected to change much with
frequency (unless the ISM electron temperature varies significantly
between locations). We can therefore attempt to trace the free-free
emission with a template. Here, we employ the estimate for the
free-free emission at $23 \, \text{GHz}$ provided by the WMAP team
using a maximum entropy method (MEM)~\cite{Gold:2008kp} and their
adopted frequency spectral index of $-2.15$.

We wish to emphasise however that the MEM free-free template is not an
unbiased tracer of thermal bremsstrahlung. First, it was derived by
separating the observed $23 \, \text{GHz}$ sky map into synchrotron,
free-free and thermal dust emission and therefore any error in the
synchrotron extrapolation translates into errors on the MEM free-free
template. Moreover, spinning dust emission was ignored all together
such that the MEM templates potentially compensate for this. The
template therefore likely overestimates the real free-free
emission. Alternatives to the MEM template, e.g. the WMAP MCMC model,
have been shown however to be even less reliable~\cite{Jaffe:2011qw}
and over- (under-)estimate free-free emission when excluding
(including) a spinning dust component. The MEM free-free template on
the other hand has a rough correlation with maps of radio
recombination lines so we adopt it here but with the above
reservations.

\subsection{Unsubtracted point sources}
\label{sec:UnsubtractedPS}

It is possible that the point source subtractions performed on the
\HaslamASS{} have not been able to subtract all the sources present in
the sky map. Relatively faint sources, and even relatively strong
sources in the Galactic disk, might have been missed. Here, we model
their APS by shot noise, as is expected for an unclustered population
of point-like objects (see~\cite{LaPorta:2008ag}), and fix the
normalisation such that the observed APS is matched at the highest $l$
accessible, i.e. $l \sim 200$. We find that 
$\mathcal{C}_l \sim 2 \times 10^{-2} \, \text{K}^2$ allows for a good fit at 
large $l$. By comparing the APS of the
unsubtracted with that of the (partly) subtracted \HaslamASS{}, we
have confirmed that this indeed describe the APS of the point
sources rather well.

\subsection{Shells of old SNRs}
\label{sec:shells}

We will see in the next section that when the APS is modelled by
adding contributions from both large- and small-scale variations of
synchrotron emissivity, as well as free-free emission and unsubtracted
point sources, there is still a large deficit in power with respect to
the observed APS on intermediate scales ($10 \lesssim l \lesssim
100$).  This cannot be explained either by an increase in free-free
emission, nor the presence of unsubtracted point sources. However, a
class of extended sources that would contribute to the APS on such
scales is visible already in sky maps from radio to microwaves (and
even in gamma-rays): the so-called `radio
loops'~\cite{Berkhuijsen:1971aa}. These huge ring-like structures,
that span up to $60^{\circ}$ in angular radius, are believed to be the
large --- up to $\sim 200 \, \text{pc}$ in radius --- shells of old
SNRs (see~\cite{Salter:1983zz,Wolleben:2007pq} for detailed studies of
Loop I, also called the North Polar Spur).

Towards the end of its lifetime, a SNR is in its radiative phase with
a low expansion velocity of $\mathcal{O}(100) \, \text{km} \,
\text{s}^{-1}$ such that particle acceleration is no longer expected
to take place (see \cite{Pohl:2008pq}). However, as the expansion is
now in the pressure-driven `snow plough' phase \cite{McKee:1977dz},
the shock compression factor $\eta$ can get very large leading to a
strong enhancement of the ISM magnetic field upon crossing the
shock. In the ISM there will be a probability distribution of $\eta$,
as was computed \cite{Sarkar:1982zz} for the McKee-Ostriker model,
however for simplicity we adopt a median value $\eta = 6$ which is
conservative. This has two important
consequences~\cite{1962MNRAS.124..125V,Sarkar:1982zz}: First, by
betatron acceleration the energy of electrons increases by a factor
$\sim \sqrt{\eta}$ upon crossing the shock; as the electron spectrum
is a steeply falling power law, this increases the emissivity for a
fixed frequency. Second, due to the compression of the magnetic field
by the same factor $\eta$, the emissivity at fixed frequency is now
sourced by electrons which are a factor $\eta$ \emph{lower} in energy
--- since these are more numerous, this further increasing the
emissivity. Hence, a compressed shell easily outshines the emissivity
of the surrounding ISM.

Given at set of shells at positions $\{ \vec{r}_i \}$, the total
synchrotron flux $J_{\text{shells}}(\nu; \ell, b)$ is the sum over all
individual shells $i$, and so are the $a_{lm}$:
\begin{equation}
J_{\text{shells}}(\nu, \ell, b) = \sum_i J_{\text{shell} \, i} (\nu,
\ell, b) \quad \Rightarrow \quad a_{lm}^{\text{shells}} = \sum_i
a_{lm}^{\text{shell} \, i} \, .
\end{equation}

We approximate the compressed medium of an old SNR as a spherical
shell of inner (outer) radius $R_1$ ($R_2$) with a constant
compression factor $\eta$. Then the flux from a single shell $i$
centred at at $\vec{r}_i$ factorises into a frequency-dependent
specific emissivity $\varepsilon_i(\nu)$ and an angular LoS integral
$g_i(l,b)$:
\begin{align}
J_{\text{shell }i} (\nu, \ell, b) &= \underbrace{\int \dd E \,
  n_{\text{acc'd}} \left(E; \vec{r}_i \right) \varepsilon(\nu, \eta
  \vec{B}(\vec{r}_i), E)}_{\varepsilon_i(\nu)} \underbrace{\int \dd s
  \, \rho \left( \left| \vec{r}(s,\theta,\phi) - \vec{r}_i \right|
  \right)}_{g_i(l,b)} \, ,
\label{eqn:Jtot}
\end{align}
where $\rho(r) = 1$ for $R_1 \leq r \leq R_2$ and $0$ otherwise. Here,
we have approximated both the betatron-accelerated electron spectrum
$n_{E; \text{acc'd}}(\vec{r})$ and magnetic field $\vec{B}(\vec{r})$
in the shell by their values at its centre $r_i$,
i.e. $n_{\text{acc'd}} \left(E; \vec{r}_i \right)$ and
$\vec{B}(\vec{r}_i)$.

\subsubsection{Specific emissivity of a shell}

Consider an electron with pitch angle $\theta_0$, i.e. momentum
$p_{\perp} = p \sin \theta_0$ ($p_{\parallel} = p \cos \theta_0$)
perpendicular (parallel) to the magnetic field. Due to efficient
compression of the GMF, the magnetic field in the shell will be mostly
aligned with the shock, that is tangential to the shell. If the shock
width is larger than the gyroradius of the (GeV) electron, the
time-scale over which the magnetic field $B$ changes is larger than a
gyration period, and therefore the first adiabatic invariant
$p_{\perp}^2/B$ is conserved. Upon crossing the shock with compression
factor $\eta$, the electron therefore gains a factor $\sqrt{\eta}$ in
$p_{\perp}$, while $p_{\parallel}$ remains constant. An initially
isotropic distribution $n(p)$ therefore becomes anisotropic:
\begin{equation}
n(p) \rightarrow n_{\text{acc'd}} (p,\theta) = n \left(
\frac{p}{\sqrt{1 + (\eta-1) \sin^2 \theta_0}} \right) \, .
\end{equation}
where the new pitch angle $\theta$ is a function of the initial pitch
angle $\theta_0$ (see \cite{1962MNRAS.124..125V} for
details). Following \cite{Sarkar:1982zz} we assume that this
distribution gets quickly isotropised by pitch-angle scattering on
Alfv\'en waves behind the shock. The distribution then needs to be
averaged over the new pitch angle $\theta$, however, an order of
magnitude approximation is given by:
\begin{equation}
n_{\text{acc'd}} (p,\theta) \approx \frac{2}{\sqrt{\eta}} n \left( p
\sqrt{\frac{3}{2 \eta}} \right) \, ,
\end{equation}
assuming that only the two perpendicular components of the momentum
change. We note that the diffusive escape of electrons from the shells
can be neglected as the magnetic field at the edge is tangential and
perpendicular diffusion is much suppressed.

The emissivity $\hat{\varepsilon}_{\text{acc'd}}$ at frequency $\nu$
of the compressed shell can now be simply expressed in terms of the
nominal emissivity $\hat{\varepsilon}$ of the (unaccelerated)
electrons in the (uncompressed) magnetic field, but at a
\emph{different} frequency:
\begin{align}
\hat{\varepsilon}_{\text{acc'd}} (\nu,\eta B) &= \int \dd E \,
\frac{2}{\sqrt{\eta}} n \left( E \sqrt{\frac{3}{2 \eta} } \right)
\varepsilon (\nu, \eta B, E) \\ &= \sqrt{\frac{8}{3}} \int \dd E' \, n
\left( E' \right) \varepsilon \left( \nu, \eta B, \sqrt{\frac{2}{3}
  \eta} E' \right) \\ &= \sqrt{\frac{8}{3}} \eta \int \dd E' \, n
\left( E' \right) \varepsilon \left( \frac{3 \nu}{2 \eta^2 }, B, E'
\right) \\ &= \sqrt{\frac{8}{3}} \eta \,
\hat{\varepsilon}_{\text{ISM}} \left( \frac{3 \nu}{2 \eta^2 }, B
\right) \, ,
\label{eqn:FrequencyPart}
\end{align}
where we have used the scaling of the specific synchrotron emissivity
(see eq.~\ref{eqn:MonoenergeticPower}).

\subsubsection{Line-of-sight integral over a shell}

\begin{figure}[tbh]
\begin{minipage}[t]{0.4\linewidth}
\centering \includegraphics[width=\columnwidth]{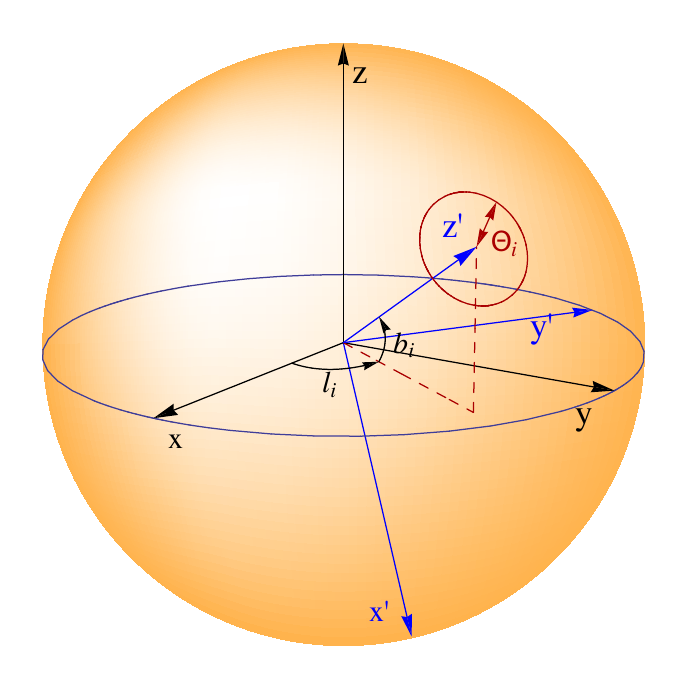}
\caption{Coordinate systems used for the computation of the APS of the
  shells. The black, unprimed coordinate system $O$ is heliocentric
  with the Galactic centre at $(x,y,z) = (8.5,0,0) \, \text{kpc}$, and
  the Galactic north pole in positive $z$-direction. The blue, primed
  coordinate system $O'$ is centred on the direction of the centre of
  a shell and can be obtained by a rotation of $\ell_i$ around the $z$
  axis and a rotation of $\pi/2 - b_i$ around the $y'$-axis.}
\label{fig:CoordinateSystems}
\end{minipage}
\hspace{0.05\textwidth}
\begin{minipage}[t]{0.53\linewidth}
\centering
\includegraphics[width=\columnwidth]{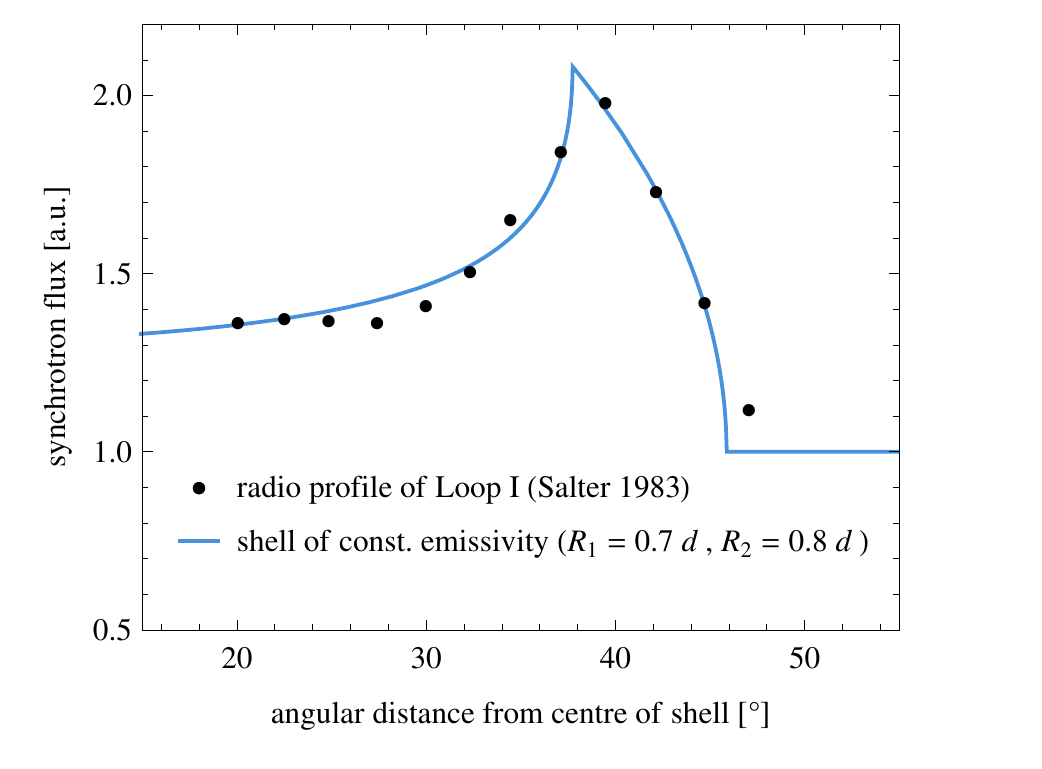}
\caption{Angular profile of a shell of an old SNR. The data points
  show the (frequency averaged) angular radio profile as measured in
  Loop I~\cite{Salter:1983zz}. The solid line is the profile of a
  shell of constant volume emissivity with inner radius $R_1 = 0.7 \,
  d$ and outer radius $R_2 = 0.8 \, d$.}
\label{fig:profile}
\end{minipage}
\end{figure}

The projection of the shell onto the celestial sphere is a
limb-brightened disk and for a given distance $d_i$, the flux depends
on the angular distance $\theta' = \arccos{ z' }$ from the projected
centre of the sphere only (see Figs.~\ref{fig:CoordinateSystems}
and~\ref{fig:profile}):
\begin{align}
g_i( z' ) = 2 \left \{
\begin{array}{llll}
\sqrt{ R_2^2 - d_i^2 (1 - z'^2) } - \sqrt{ R_1^2 - d_i^2 (1 - z'^2) }
& \quad \text{for} \quad & \sqrt{d_i^2 (1 - z'^2)} \leq R_1 \, ,
\\ \sqrt{ R_2^2 - d_i^2 (1 - z'^2) } & \quad \text{for} \quad R_1 < &
\sqrt{d_i^2 (1 - z'^2)} \leq R_2 \, , \\ 0 & \quad \text{otherwise} \,
.
\end{array}
\right.
\label{eqn:profile}
\end{align}

We start by considering the $a'^i_{lm}$ of an individual shell $i$ in
a coordinate system $O'$ with $z'$-axis in the direction of the centre
of the shell (see Fig.~\ref{fig:CoordinateSystems}),
\begin{align}
a'^i_{lm} &= \varepsilon_i( \nu )\int \dd \Omega'
Y_{lm}^*(\theta',\phi') \, g(z') \\ &= \varepsilon_i( \nu )\int_{-1}^1
\dd z' \int_0^{2 \pi} \dd \phi' \sqrt{\frac{2l+1}{4 \pi}}
\sqrt{\frac{(l-m)!}{(l+m)!}} P_l^m(z') \ee^{i m \phi'} g(z') \\ &=
\left \{
\begin{array}{llll}
\varepsilon_i( \nu )\sqrt{\frac{2l+1}{4 \pi}} \int_{-1}^1 \dd z'
P_l(z') g(z') & \quad \text{for} \quad & m = 0 \, , \\ 0 & \quad
\text{otherwise} \, .
\end{array}
\right.
\end{align}

The spherical harmonics coefficients $a^i_{lm}$ in Galactic
coordinates $O$ can be computed using the Wigner D-matrices $D_l^{m'
  m}(-\ell_i , \pi/2 - b_i, 0)$,
\begin{align}
a^i_{lm} &= \varepsilon_i( \nu )\sum_{m' = -l}^l D_l^{m' m} \left(
-\ell_i, \frac{\pi}{2} - b_i, 0 \right) a'^i_{lm'} \\ &=
\varepsilon_i( \nu )D_l^{0 m} \left( -\ell_i, \frac{\pi}{2} - b_i, 0
\right) \sqrt{\frac{2l+1}{4 \pi}} \int_{-1}^1 \dd z' P_l(z') g(z') \\
&= \varepsilon_i( \nu )Y^*_{lm} \left( \frac{\pi}{2} - b_i, -\ell_i
\right) \int_{-1}^1 \dd z' P_l(z') g(z') \, .
\label{eqn:almi}
\end{align}

The integral in the last line can be written:
\begin{align}
\int_{-1}^1 \dd z' P_l(z') g(z')
&= d J_l( \sqrt{1-(R_2/d)^2} ) - d J_l( \sqrt{1-(R_1/d)^2} ) \, ,
\end{align}
where
\begin{equation}
J_l(a) = \int_a^1 \dd z' P_l(z') \sqrt{z^2-a^2}
\end{equation}
can be solved analytically in terms of Gauss hypergeometric functions
${}_2 F_1$ (cf. App.~\ref{sec:integral}).

\subsubsection{The Galactic population of radio loops}

Only four of the radio loops are visible in the sky and can be
modelled directly. We fix their radius to $200 \, \text{pc}$ which is
consistent both with calculations of the evolution of SNRs
\cite{McKee:1977dz,Bandiera:2004ii} and estimates for Loop I
\cite{Salter:1983zz,Wolleben:2007pq}. We have adopted the positions
and angular sizes as found originally~\cite{Berkhuijsen:1971aa} and
computed the distances, see Table~\ref{tbl:LocalShells}.

\begin{table}[!tbh]
\centering
\begin{tabular}{| l | l | l | l | |l |}
\hline
Object & $l$ (centre) & $b$ (centre) & diameter & distance \\
\hline
Loop I & $329^{\circ}$ & $17.5^{\circ}$ & $116^{\circ}$ & $240 \, \text{pc}$ \\
Loop II & $100^{\circ}$ & $-32.5^{\circ}$ & $91^{\circ}$ & $280 \, \text{pc}$ \\
Loop III & $124^{\circ}$ & $15.5^{\circ}$ & $65^{\circ}$ & $370 \, \text{pc}$ \\
Loop IV & $315^{\circ}$ & $48.5^{\circ}$ & $39.5^{\circ}$ & $590 \, \text{pc}$ \\
\hline
\end{tabular}
\caption{Angular positions, diameters and resulting physical distances
  of the local radio loops for an assumed physical radius of $200 \,
  \text{pc}$~\cite{Berkhuijsen:1971aa}.}
\label{tbl:LocalShells}
\end{table}

However, the observable loops constitute most likely only a small
fraction of the total number of old SNRs in the Galaxy. Adopting a SN
rate of $1-3 \times 10^{-2} \, \text{yr}^{-1}$ and estimating the
duration of the radiative phase to be between $10^4-10^5 \,
\text{yr}$, we can expect up to \emph{thousands} of those old SNR
shells to contribute to the synchrotron emission in radio and
microwaves. Most of them will however not be directly observable in
the sky since being further away than the 4 local loops they will
blend into the diffuse emission from the Milky Way
\cite{Sarkar:1982zz}. We therefore model them by Monte Carlo by
randomly choosing 1000 positions distributed according to the
\textit{pulsar-like} source distribution adopted by \texttt{GALPROP}
(see Sec.~\ref{sec:large-scale} and
Table~\ref{tbl:GALPROP_parameters}). As we will add the contribution
from this random set to that from the known local loops, we have
excised from the random set all those shells that are closer than $500
\, \text{pc}$ in order not to double-count the local shells.

We fix the radius of all shells to $200 \, \text{pc}$ and calculate
their spherical harmonics expansion coefficients $a_{lm}^i$ from
eqs.~\ref{eqn:FrequencyPart} and \ref{eqn:almi}. Finally, we sum up
the spherical harmonics expansion coefficients to
$a_{lm}^{\text{shells}}$ and compute their angular power spectrum
$\mathcal{C}_l^{\text{shells}}$.

\section{Results}

\subsection{APS from conventional model}

\begin{figure}[tbh]
\centering
\includegraphics[width=\textwidth]{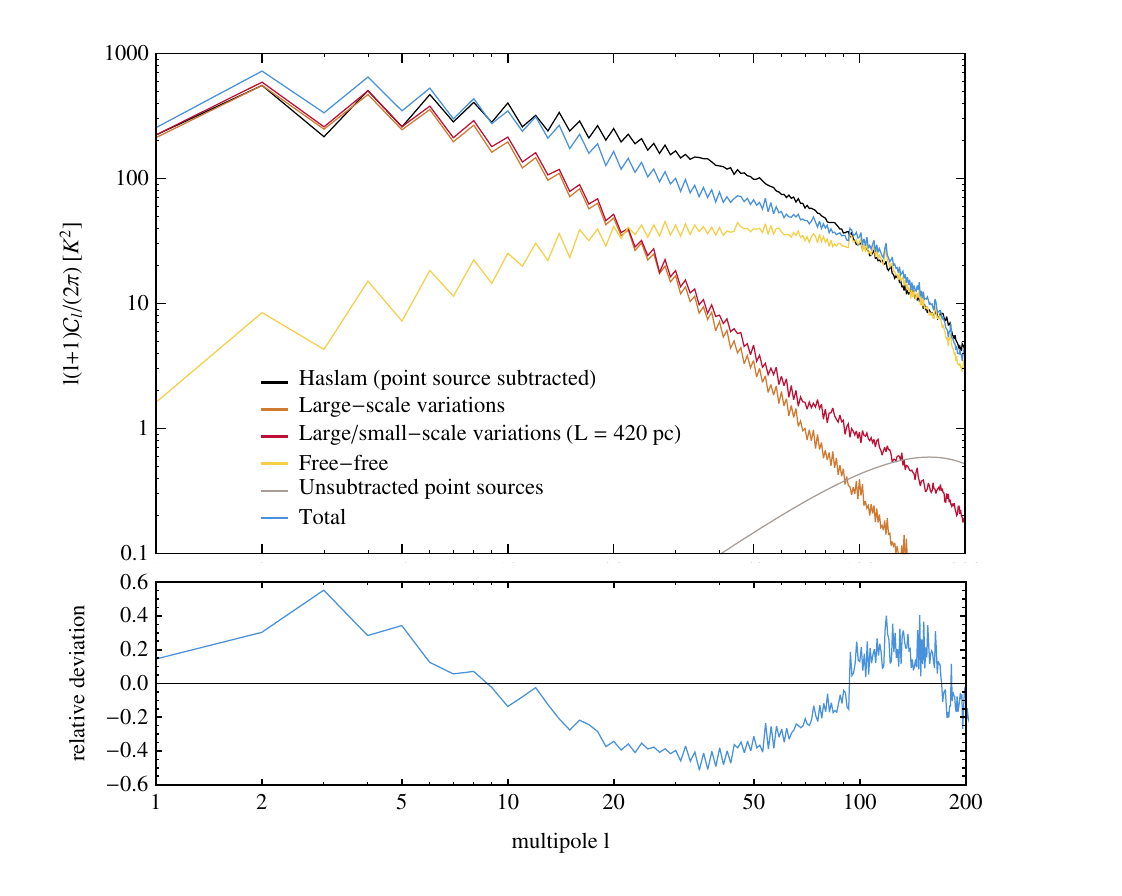}
\caption{{\bf Top panel:} The APS of the synchrotron sky computed by
  taking into account the large-scale and small-scale variations of
  emissivity (orange line), free-free emission (purple line) and
  unsubtracted sources (grey line). The sum of these components (blue
  line) is compared with the APS from the \HaslamASS{} (black
  line). {\bf Bottom panel:} The relative residual of the modelled APS
  with respect to the observed one.}
\label{fig:APS_wo_shells}
\end{figure}

In Fig.~\ref{fig:APS_wo_shells} we show our modelling of the angular
power spectrum resulting from both large-scale and small-scale
variations of synchrotron emissivity as well as free-free emission and
unsubtracted point sources and compare it to the observed APS of the
\HaslamASS{}. For all modelled components we have taken into account 
the exponential cut-off due to the beam window function.

We start the discussion by computing the APS from the variation of the
synchrotron emissivity on large scales only (orange line).  We have
normalised this component by matching the dipole ($l=1$) which
required decreasing the intensity by $28 \, \%$.  We note that most of
the power is indeed contained on large scales ($l \leq 10$) and the
APS falls off quite rapidly for higher $l$, as is indeed expected
given that all the ingredients (source distribution, transport
parameters and magnetic fields) vary only on kpc scales. The APS
computed with the parameters of~\cite{Strong:2010pr,Strong:2011wd}
does show the expected even-odd multipole structure but the difference
between the power in even and odd multipoles is too small. This
signals that the distribution in longitude is not extended enough; if
it were more extended this would bring us closer to the symmetric case
and would suppress power in the even multipoles. In
Fig.~\ref{fig:APS_wo_shells} and henceforth we have therefore adopted
$\rho = 100 \, \text{kpc}$.

Next we compare the contribution of the ISM synchrotron emissivity,
with (red line) and without (orange line) the small-scale
turbulence. We note that while the turbulent APS indeed shows a power
law behaviour $\mathcal{C}_l \propto l$ for $l \gtrsim 50$, the
additional power due to turbulence is relatively small. This
corroborates our arguments from \S~\ref{sec:small-scale} about the
relative power in the monopole and in higher multipoles and
demonstrates that the large-scale behaviour of the measured APS cannot
be due to ISM turbulence.

Now adding the contribution from free-free emission (yellow line) and
the $\mathcal{C}_l$ due to unsubtracted sources (grey line, almost
negligible below $l \sim 100$), we arrive at the total APS (blue
line). Comparing this to the measured APS, it is evident that ISM
synchrotron and free-free emission alone can give a reasonable fit
only for the dipole and quadrupole and around $l \sim 100$. At all
other $l$, and in particular at intermediate multipoles, the model
\emph{underproduces} the power observed in the \HaslamASS{}.

\subsection{APS adding SNR shells}

\begin{figure}[tbh]
\centering
\includegraphics[width=\textwidth]{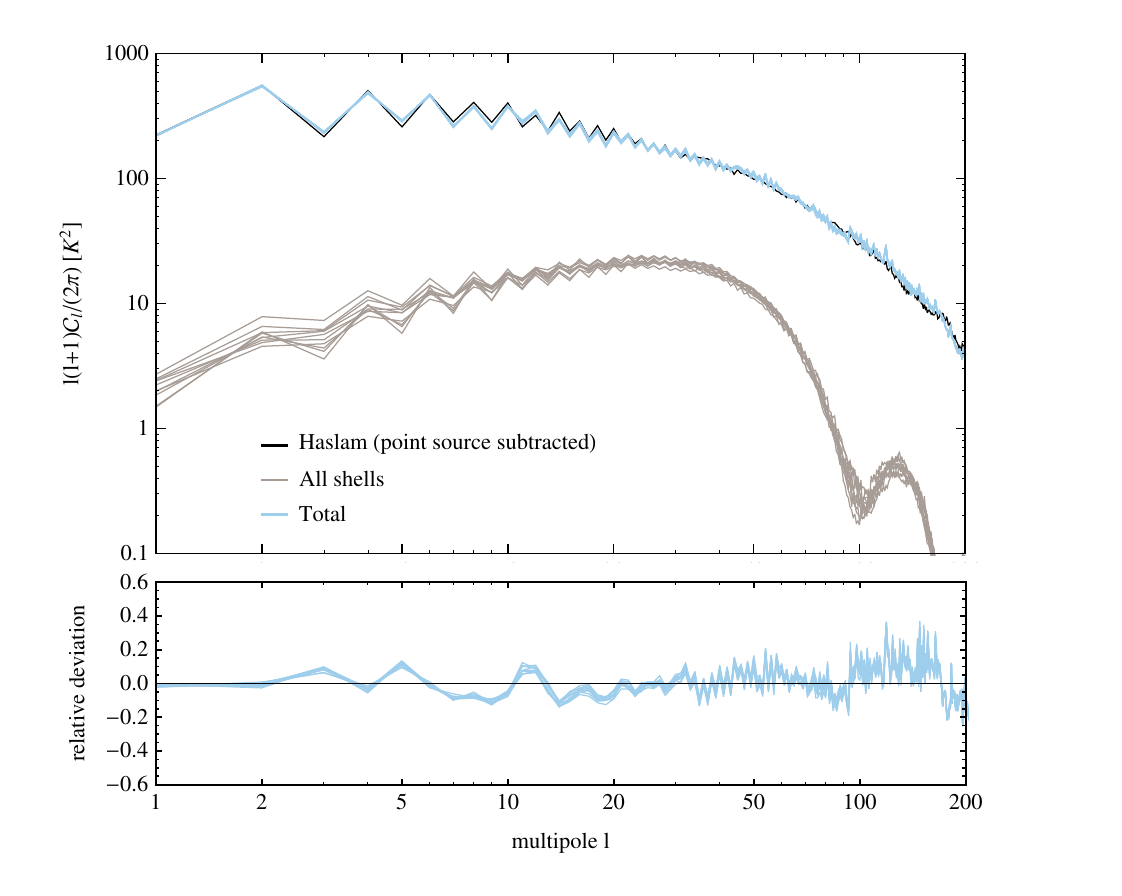}
\caption{{\bf Top panel:} The APS of the synchrotron sky computed by
  summing the large-scale and small-scale variations of emissivity,
  free-free emission, unsubtracted sources as well as the contribution
  from different configurations of shells of old SNRs (also shown
  separately). {\bf Bottom panel:} The relative residual of the
  modelled APS with respect to the observed one.}
\label{fig:APS_w_shells_ensemble}
\end{figure}

Figure~\ref{fig:APS_w_shells_ensemble} shows the modelled APS as
before, but now including the contribution from the four local plus
the $\sim 1000$ shells distributed in the Galaxy, for 10 different
realisations of this ensemble. The APS for the shells alone is shown
to illustrate that these contribute on large and intermediate scale,
i.e. at $l \lesssim 50$ which is where we previously noted a
deficit. (As the shells were added to the \texttt{GALPROP} model, we
needed to further renormalise the contribution from the ISM emissivity
to about $70 \, \%$.) The agreement with the measured APS is now much
improved with residuals at most $20 \,\%$. The differences between the
Monte Carlo realisations of the Galactic distribution of old SNRs is
relatively small which illustrates our point that the APS is
particularly appropriate for modelling of Galactic foregrounds as the
influence of unknown parameters (e.g. the positions of and distances
to the shells) is rather mild.  Since the shells do not dominate the
power at any $l$, the slightly larger residuals for $l \gtrsim 100$
must be attributed to the free-free MEM map, concerning which we have
expressed some reservations already.

\begin{figure}[ht]
\centering
\includegraphics[width=\textwidth]{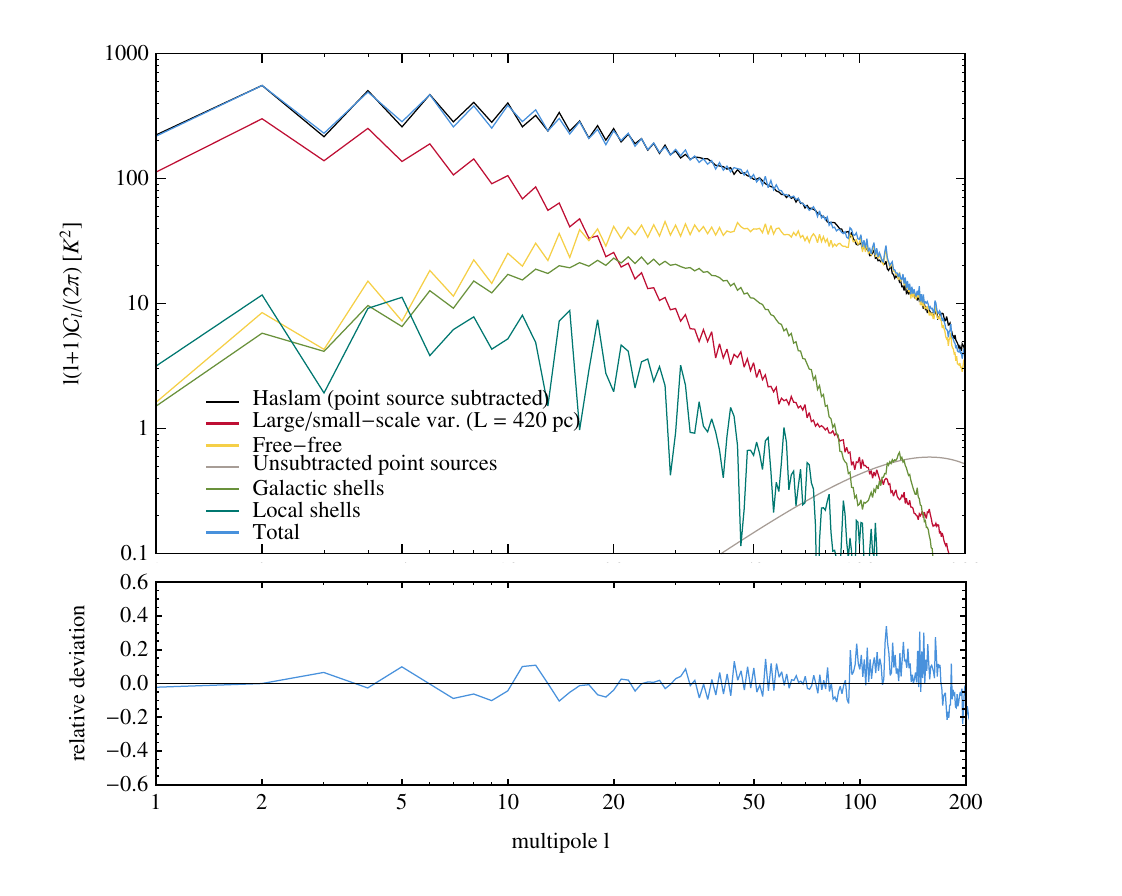}
\caption{{\bf Top panel:} The APS of the synchrotron sky obtained by
  summing the large-scale and small-scale variations of emissivity,
  free-free emission, unsubtracted sources, as well as the contribution
  from the best fit configuration of shells of old SNRs (also shown
  separately). {\bf Bottom panel:} The relative residual of the
  modelled APS with respect to the observed one.}
\label{fig:APS_w_shells_bestfit}
\end{figure}

In Fig.~\ref{fig:APS_w_shells_bestfit} we show the best fit of our
simple model to the observed APS --- this has residuals no bigger than
$10 \, \%$ up to $l \sim 200$, i.e. down to a degree.

\section{Discussion}

The reader might wonder why we attempt to develop a physical model of
the Galactic synchrotron emission in order to e.g. perform a reliable
foreground subtraction for the CMB, since it is supposedly possible to
measure CMB temperature anisotropies successfully even \emph{without}
a physical model of the foregrounds~\cite{Gold:2010fm}. This is
justified by saying that most foregrounds (in particular on small
scales, i.e. at high $l$) are contained in the Galactic disk and can
therefore be suppressed by simply masking this part of the
sky. However apart from complicating the analysis (spherical harmonics
on the cut sky are \emph{not} orthonormal, thus introducing
correlations between the $a_{lm}$ of different $l$), it is hard to
estimate by how much the masked sky maps are still contaminated by
\emph{high-latitude} Galactic foregrounds. It is quite possible that
the low multipole anomalies of the CMB are at least partly due to such
residual contamination. Even in e.g. the WMAP 9-year data, the
cosmological model fit is particularly poor at low multipoles and this
is acknowledged ``to be at least partially due to residual
foregrounds'' \cite{Bennett:2012fp}.

Second, our modelling of the unpolarised synchrotron emission is only
a first step, with the next challenge being a sound understanding of
the polarised component which will need to be subtracted off for
measurement of the CMB $B$-mode polarisation due to gravitational
waves from inflation. At 90 GHz the expected signal for a realistic
value of the tensor-to-scalar ratio $r \sim 0.01$ reaches a maximum of
$\sim10-30\%$ of the estimated foreground on degree scales, over 75\%
of the sky \cite{Dunkley:2008am}. Hence to reliably extract this will
require an understanding of the polarised foreground at the 10 \%
level. Our fit to the unpolarised emission is already accurate to $10
\,\%$ and a considerable improvement over previous studies of the
synchrotron emission in sky maps. We emphasise that this has not been
achieved by means of a detailed parameter study of existing models but
by an alternative representation --- the APS --- of the information
contained in the sky map, and comparison with the simplest possible
models, i.e. the standard \texttt{GALPROP} setup and simple physics
describing the shells of old SNRs. This illustrates the power of the
APS approach for modelling diffuse foregrounds --- applications are
certainly not constrained to radio and microwave frequencies alone.

As to the changes necessary for modelling the polarised emission, one
needs of course to model the ordered component of the GMF (which we
have neglected here), but in principle the other ingredients remain
the same. In particular if we follow the picture of old SNR shells as
compressions of the ISM together with the GMF, the magnetic fields
should be mostly tangential which should enable easy modelling of the
polarised emission from the shells. This simple picture seems to be
borne out by observations of polarised emission in microwaves from the
four local loops which show the polarisation vectors mostly radially
orientated (see Fig.~10 of~\cite{Page:2006hz}). The contribution from
thermal dust can be constrained from Planck observations at the
highest frequencies.  Furthermore, given that the (uncertain)
free-free emission is unpolarised, we are confident that our approach
will allow a similar (if not better) level of precision in
polarisation.

\begin{figure}[bth]
\begin{tabular}{cc}
\includegraphics[width=0.48\textwidth]{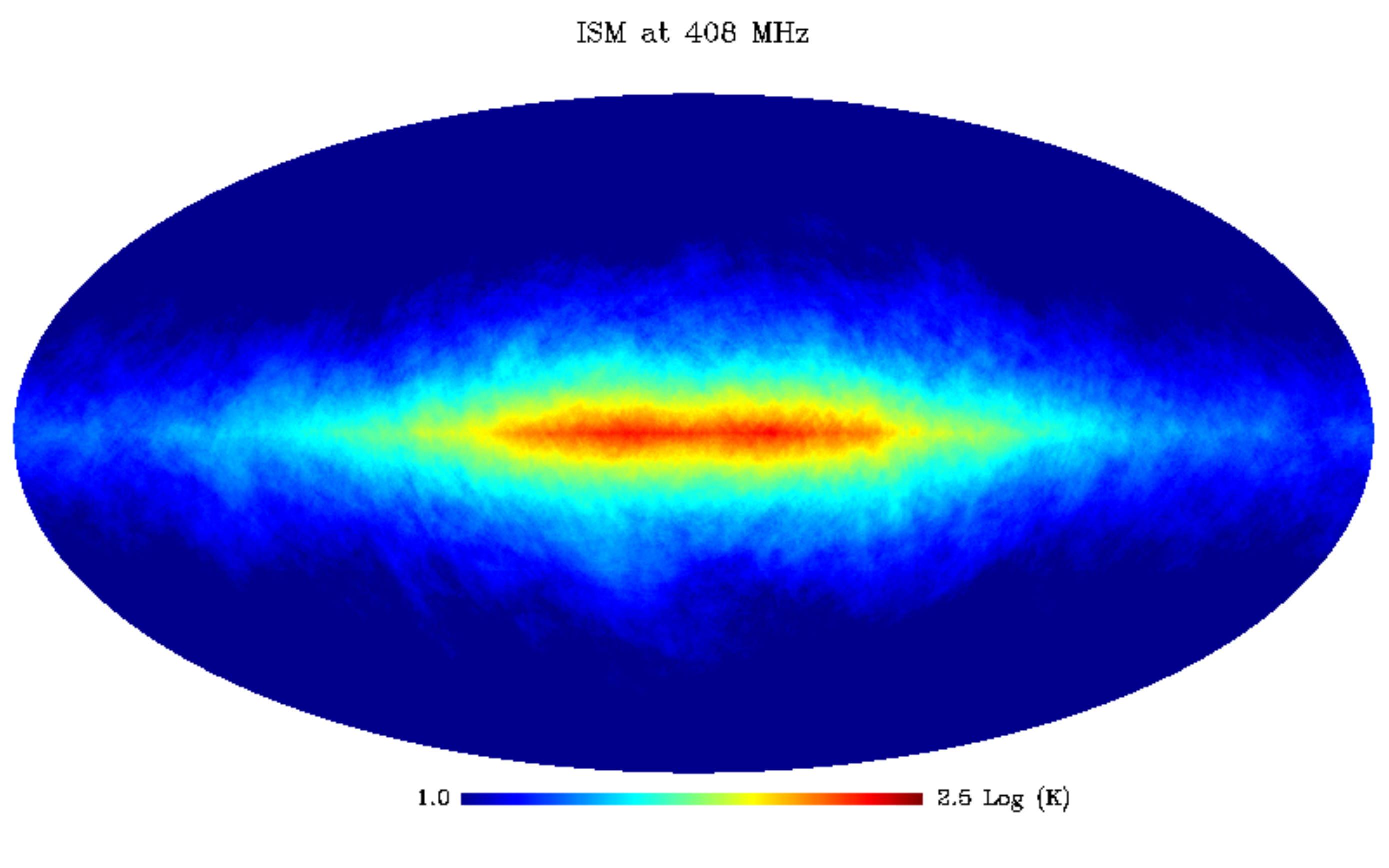} &
\includegraphics[width=0.48\textwidth]{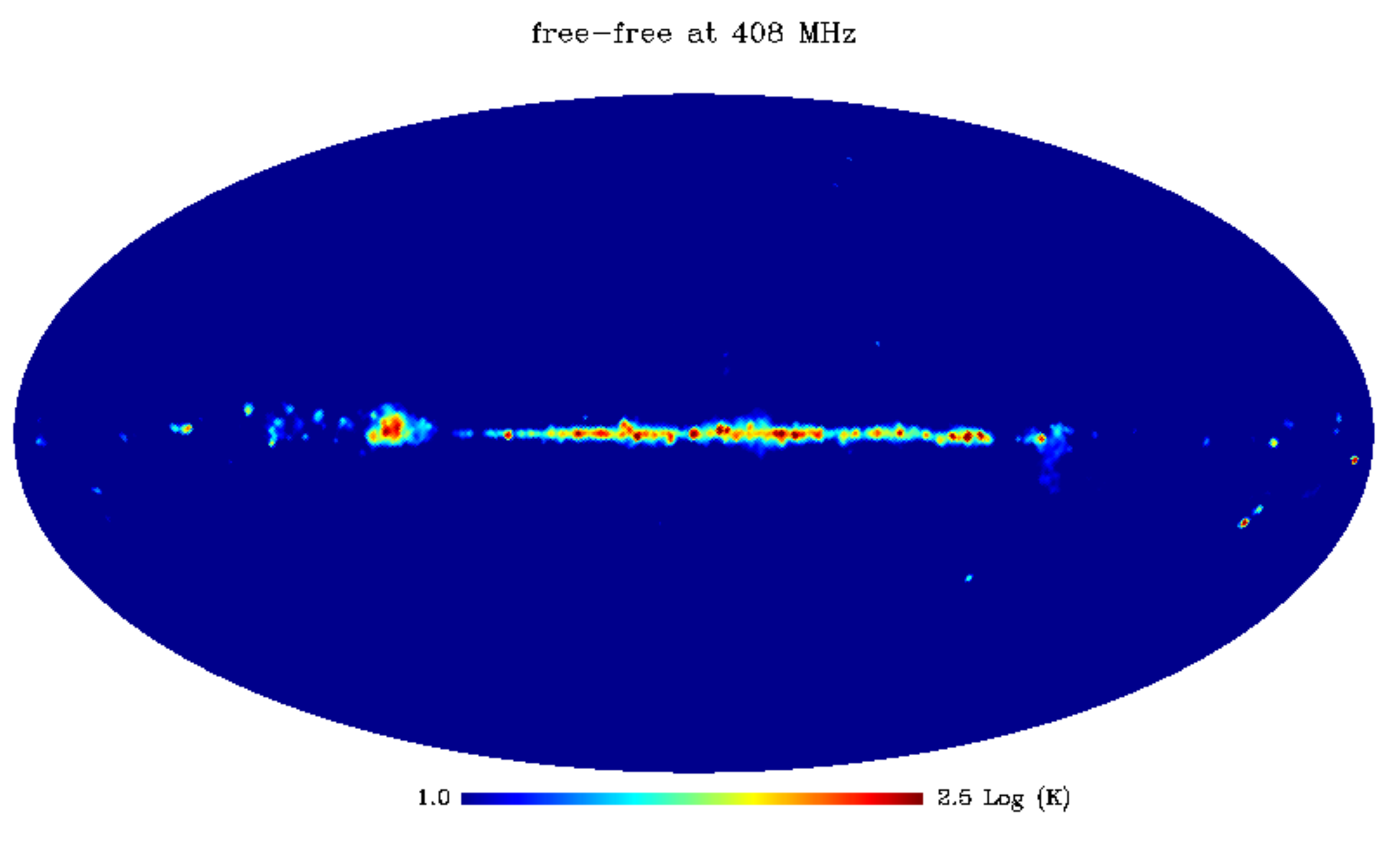} \\
\includegraphics[width=0.48\textwidth]{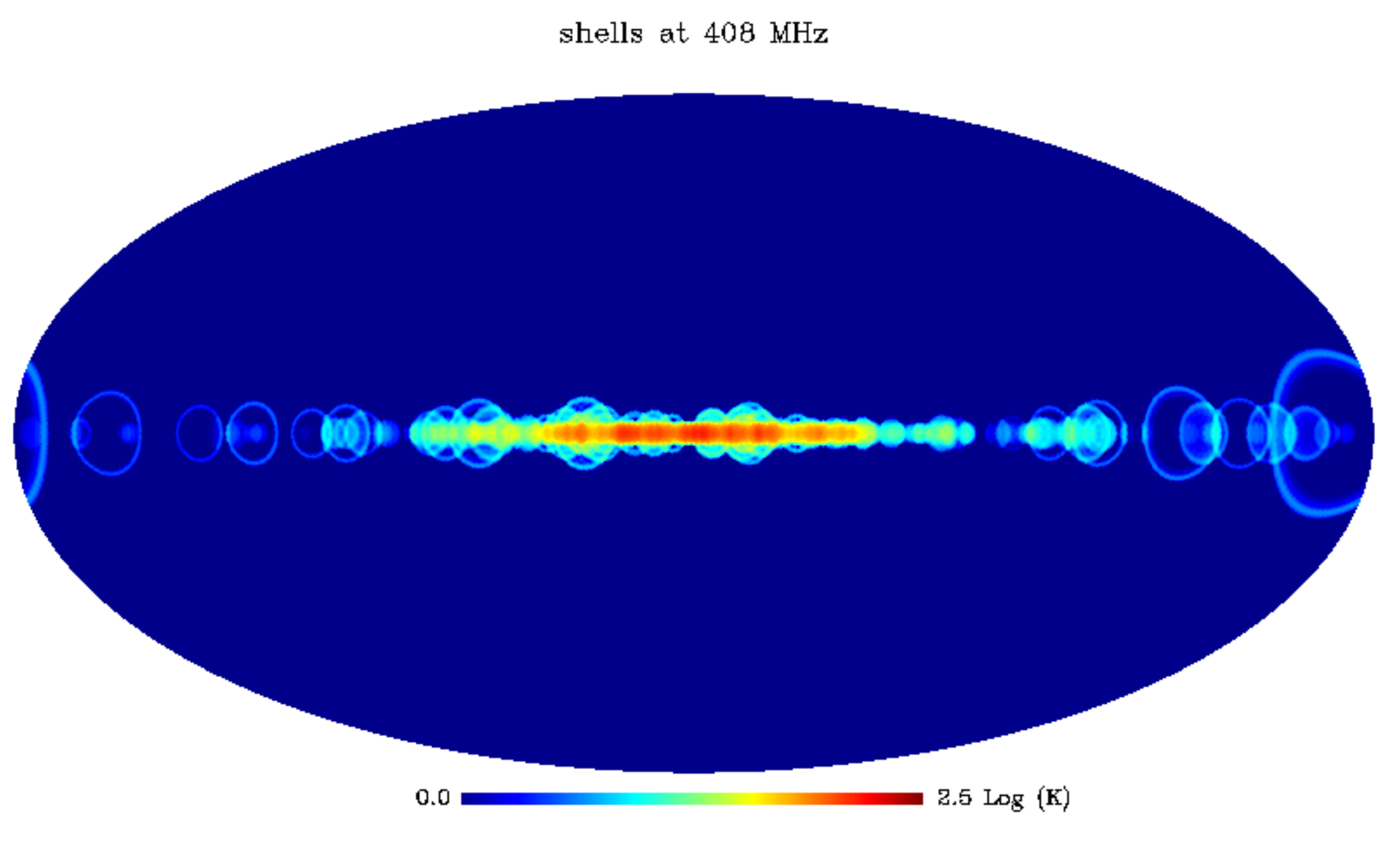} &
\includegraphics[width=0.48\textwidth]{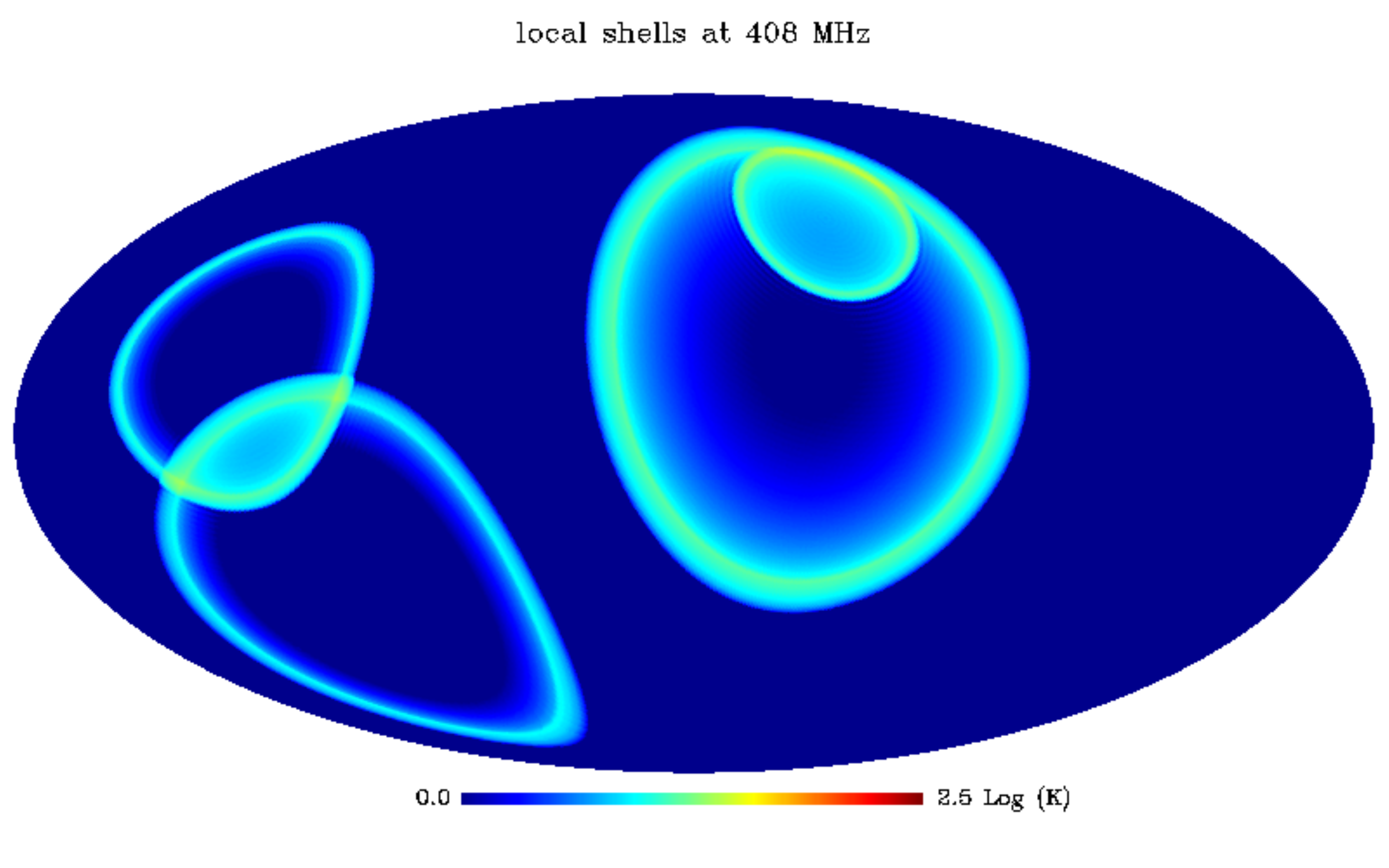} \\
\includegraphics[width=0.48\textwidth]{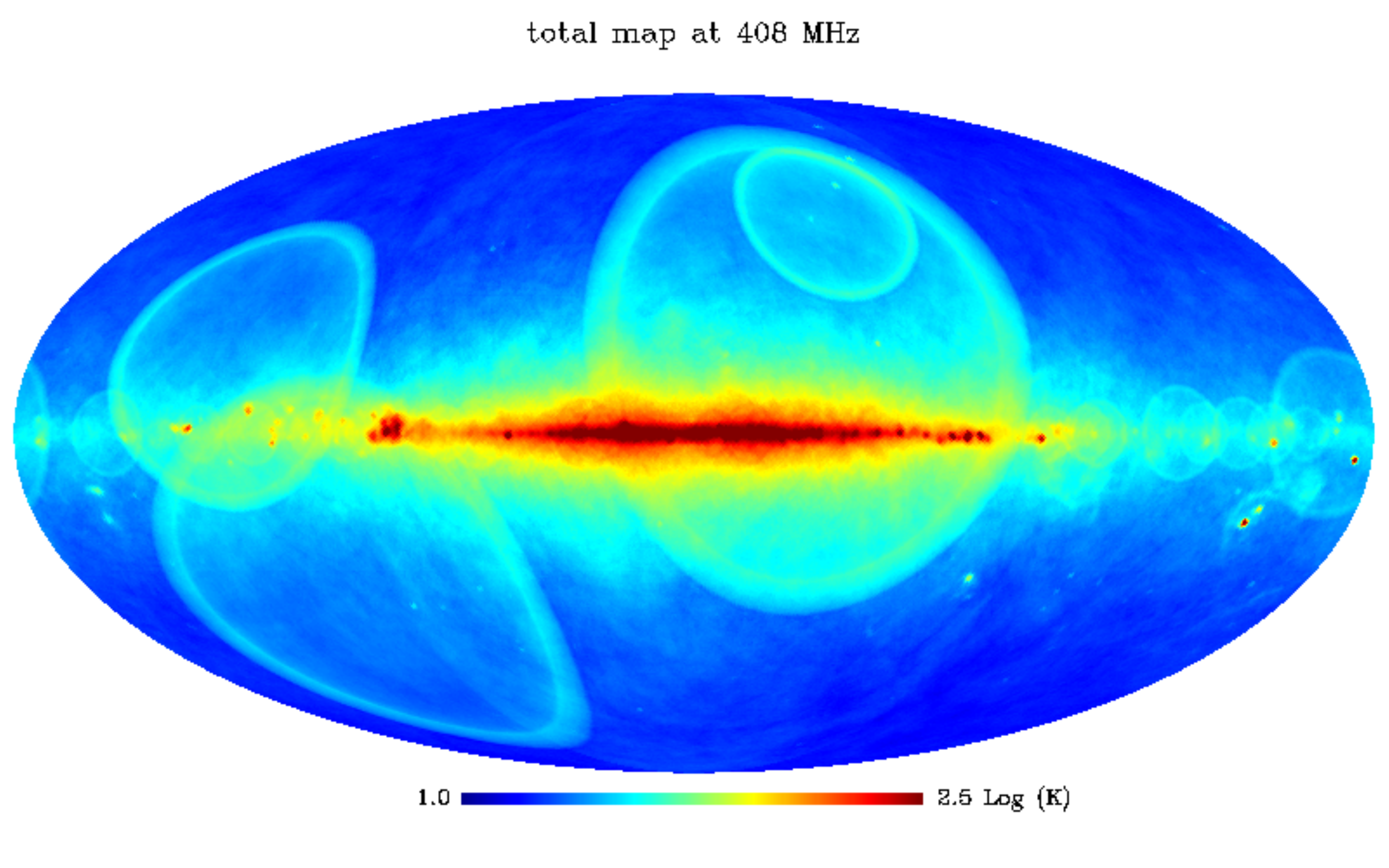} &
\includegraphics[width=0.48\textwidth]{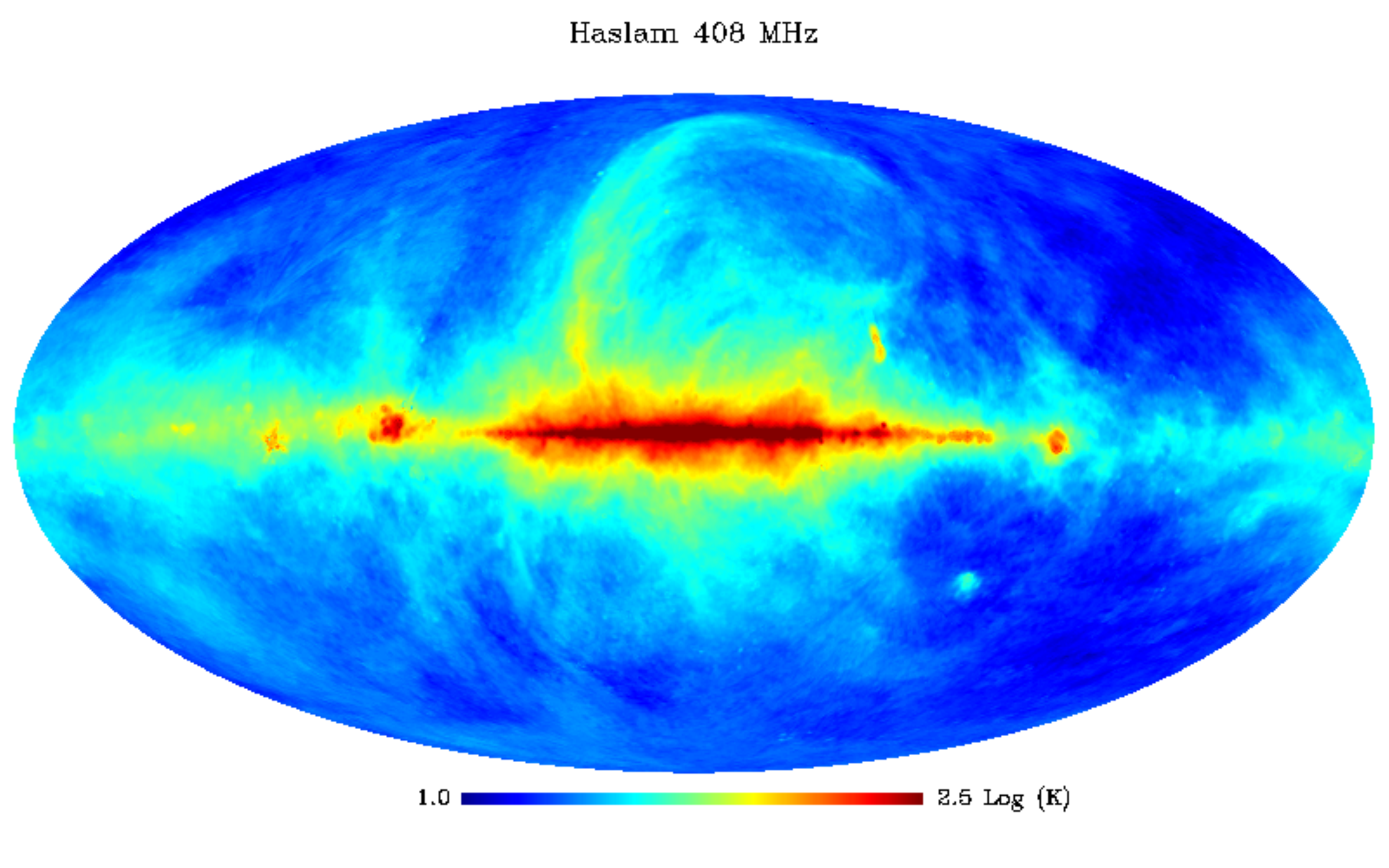} \\
\end{tabular}
\caption{ {\bf Top left:} ISM synchrotron emission on large (modelled
  by \texttt{GALPROP}) and small scales.  {\bf Top right:} Free-free
  emission, extrapolated from the WMAP MEM free-free map.  {\bf Middle
    left:} Synchrotron emission from the Galactic population of shells
  of old SNRs.  {\bf Middle right:} Synchrotron emission from the four
  local shells of old SNRs, viz. Loop I - IV.  {\bf Bottom left:} The
  sum of synchrotron emission on large and small scales, free-free
  emission, Galactic population of shells and local shells.  {\bf
    Bottom right:} The \HaslamASS{}.  }
\label{fig:components}
\end{figure}

Although we have stressed the validity of modelling of the APS alone,
for context we also show in Fig.~\ref{fig:components} sky maps of the
different components contributing to the diffuse emission at
\Haslam{}. With the inclusion of turbulence up to an outer scale of
$400 \, \text{pc}$, the diffuse emission of the ISM (top left panel)
looks very similar, in particular the level of ``patchiness''. However
it appears that the emission in the Galactic plane (lower right panel)
is not as peaked as is observed. (To some extent this could have been
anticipated as the \texttt{GALPROP} model is not tuned to match radio
angular profiles.) The Galactic population of SNR shells (middle left
panel) does contribute mainly in the plane, since most of them are far
away, and brings the latitudinal profile into agreement with the
observed one. Note that there are still a couple of shells at
distances $\lesssim 1 \, \text{kpc}$ that lead to visible loops in the
sky map. The most obvious difference to the sky map for the local
shells, i.e. those with distances $\lesssim 500 \, \text{pc}$ (middle
left panel) is, apart from the larger angular size, their
\emph{offset} from the Galactic plane. (We have constrained the
centres of the non-local shells in our MC simulation to lie in the
Galactic plane but loosening this constraint does not change the APS
much.) Finally, the sum of all components (bottom left panel) shows a
remarkable resemblance to the observed sky map (bottom right
panel). Note that the observed radio loops do not possess the perfect
ring-like structures as assumed in our simplified approach in
\S~\ref{sec:shells}; the emissivity probably varies due to the
non-uniform evolution of the SNR, the overall structure of the ordered
GMF and the interaction and overlap of shells. However, the APS is
mostly sensitive to the overall angular structure so this does not
result in too big a deviation.

\section{Conclusions}

We have presented a new model for the diffuse, Galactic synchrotron
radio background that takes into account variations of the emissivity
from the scale of the Galaxy as a whole, down to the scale of
interstellar magnetic turbulence. We uncovered a lack of power on
intermediate scales when only the variation of the ISM emissivity is
taken into account (besides free-free emission and unsubtracted point
sources). We argued that this deficit is most likely cured by the
inclusion of angular correlations due to the presence of ${\cal
  O}(1000)$ old SNR shells, which are similar to the known four local
`radio loops'. Even in the simplest approach, i.e. when we fix the
free parameters using other observables, the agreement with
observations at \Haslam{} is quite remarkable. This is mainly because
the chosen observable, i.e. the APS, is particularly appropriate in
the sense that it averages out much of the uninteresting stochasticity
and highlights the dependence on the underlying physical model. We
believe that the agreement can be improved even further by calibrating
our model to (point source subtracted) all-sky maps at other
frequencies where synchrotron emission is still dominant, e.g. at
$1420
\,\text{MHz}$~\cite{1982A&AS...48..219R,1986A&AS...63..205R,2001A&A...376..861R}
and 2326 MHz~\cite{1998MNRAS.297..977J}. This will be addressed in
future work..

\section{Acknowledgements}

SS thanks the Niels Bohr International Academy, Copenhagen for 
a Visiting Professorship. PM thanks Prof Pavel
Naselsky of the `Discovery Center', Niels Bohr Institute for support
and for very useful discussions. This work was supported in part by the 
Department of Energy contract DE-AC02-76SF00515 and the KIPAC 
Kavli Fellowship made possible by The Kavli Foundation. We have 
benefitted from very helpful comments by the Referee.

\appendix

\section{The monopole compared to the total power}
\label{sec:Monopole}

\begin{figure}[tbh]
\begin{minipage}[t]{0.35\linewidth}
\centering
\includegraphics[width=0.8\textwidth]{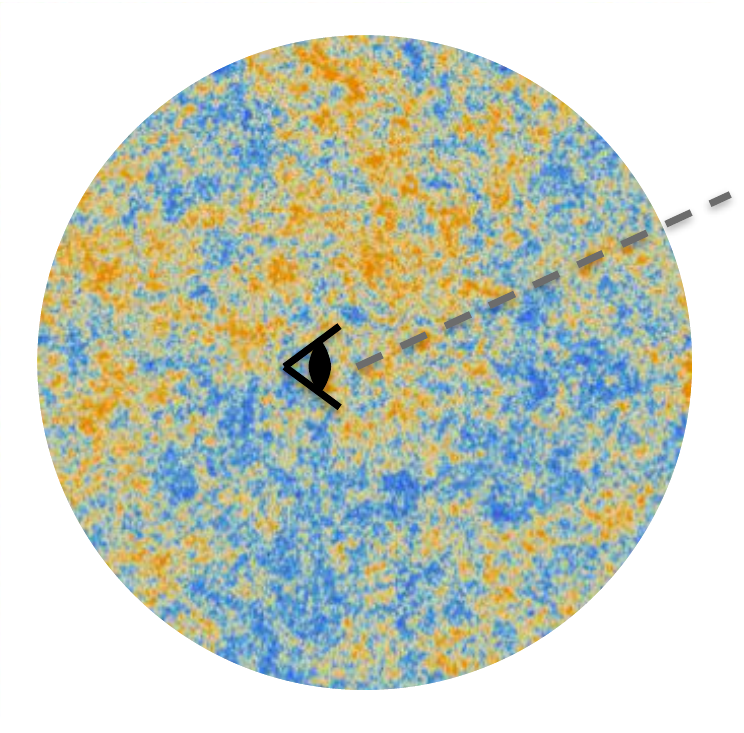}
\caption{Observer at the centre of a turbulent sphere as in the
  analytical computation for small-scale varying turbulence.}
\label{fig:StatIsoSetup}
\end{minipage}
\hspace{0.04\textwidth}
\begin{minipage}[t]{0.6\linewidth}
\centering
\includegraphics[width=0.7\textwidth]{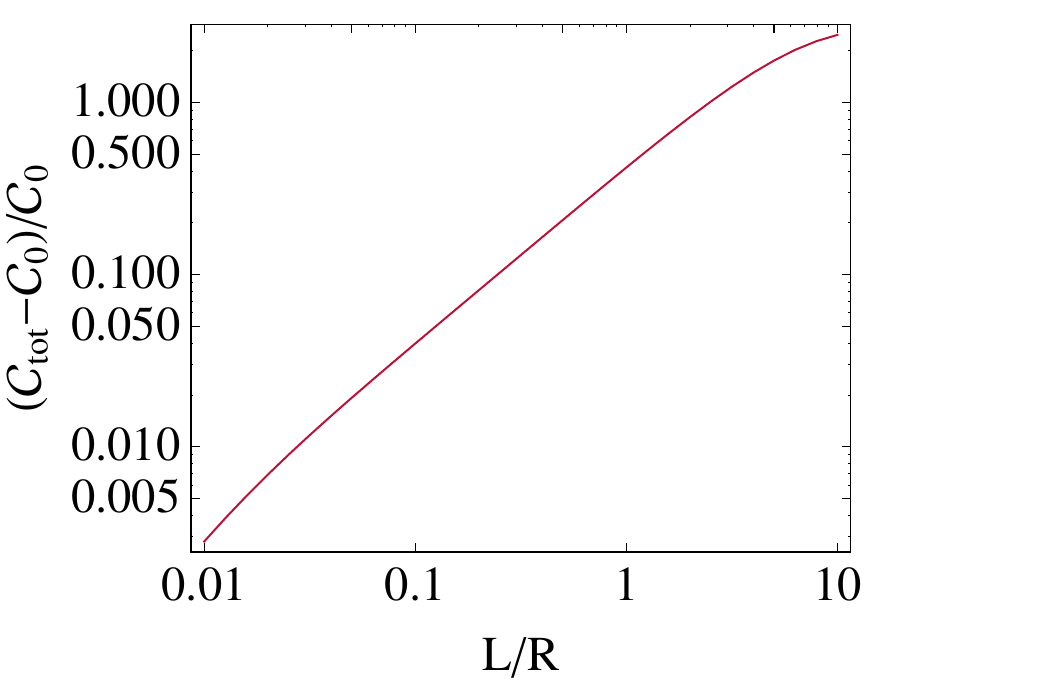}
\caption{Power in modes $l>0$ relative to the power in the monopole,
  as a function of $L/R$.}
\label{fig:monopole2total}
\end{minipage}
\end{figure}

Given a (rescaled) synchrotron skymap,
\begin{equation}
J(\theta, \phi) = \int \dd s \, w(s) B_{\perp}^2 (\vec{r}(s,\theta,\phi)) ,
\end{equation}
the monopole is proportional to the squared average of this sky map:
\begin{align}
\mathcal{C}_0 = |a_{00}|^2 = 4 \pi \left( \frac{1}{4\pi} \int \dd
\Omega \, J(\theta, \phi) \right)^2 = 4 \pi \frac{2}{3} \left(
\frac{\sqrt{\pi}}{2} R \int \dd^3 k \, \frac{\dd B^2}{\dd^3 k}
\right)^2 \, .  \label{eqn:monopole}
\end{align}
The total power is, on the other hand, defined as the square of the
map integrated over the whole sky,
\begin{equation}
\mathcal{C}_{\text{tot}} = \int \dd \Omega \, J^*(\theta, \phi)
J(\theta, \phi) = \sum_l (2l+1) \mathcal{C}_l \, ,
\end{equation}
i.e. the weighted sum of the APS. In the present case, i.e. for a
power law turbulence spectrum
\begin{equation}
\frac{\dd B^2}{\dd^3 k} = \mathcal{F}_0^2 k^{-11/3} \ee^{- k_0^2/k^2} \, ,
\end{equation}
it can be
computed via eq.~(\ref{eqn:APSsmallscale}). In
Fig.~\ref{fig:monopole2total} we show $(\mathcal{C}_{\text{tot}} -
\mathcal{C}_0 )/\mathcal{C}_0$, i.e. the power in modes $l>0$ relative
to the power in the monopole, as a function of $L/R$. This is
essentially the variance of a sky map $J(\theta, \phi)$ divided by its
mean square:
\begin{equation}
\frac{\mathcal{C}_{\text{tot}} - \mathcal{C}_0 }{\mathcal{C}_0} =
\frac{\langle J^2 \rangle - \langle J \rangle^2}{\langle J \rangle^2}
,
\end{equation}
where the angled brackets denote averages over the sky. It is apparent
that the multipoles $l>0$ only contribute to the total power of the
sky map when $L/R > 1$. However, this would mean that the outer scale
of turbulence is larger than the column depth -- a contradiction in
terms. We conclude that for realistic parameter combinations, e.g. $L
\sim 100 \, \text{pc}$ and $R > 1 \, \text{kpc}$ the monopole always
dominates the total power of the map. This is essentially a
consequence of the central limit theorem: as the number of turbulent
eddies that is averaged over by the LoS integration becomes large, the
variance around the average gets small.

\section{Auxiliary function $J_l(a)$}
\label{sec:integral}

We start evaluating the integral $J_l(a)$ by partial integration and
by using recurrence relations of the Legendre polynomials,
\begin{align}
J_l(a) 
&= \int_a^1 \dd z' P_l(z') \sqrt{z^2-a^2} \\
&= \left[ \frac{1}{2l+1} \left( P_{l+1}(z) - P_{l-1}(z) \right) \sqrt{z^2-a^2} \right]_a^1 \\
& -\frac{1}{2l+1} \int_a^1 \dd z \frac{z}{\sqrt{z^2-a^2}} \left( P_{l+1}(z) - P_{l-1}(z) \right) \\
&= -\frac{1}{2l+1} \int_a^1\frac{\dd z}{\sqrt{z^2-a^2}} \left( \frac{l+2}{2l+3} P_{l+2}(z) + \left( \frac{l+1}{2l+3} - \frac{l}{2l-1} \right) P_l(z) - \frac{l-1}{2l-1} P_{l-2}(z) \right) \, .
\end{align}
Integrals of this form compute as the real part of the same integrand
but with the lower boundary at $0$:
\begin{align}
\int_a^1\frac{\dd z}{\sqrt{z^2-a^2}} P_l(z) 
&= \Re \left\{ \int_0^1\frac{\dd z}{\sqrt{z^2-a^2}} P_l(z) \right\} \, .
\end{align}
The integral in the brackets can now be solved by exploiting that $\dd
(\operatorname{arsinh}\, z) / \dd z = 1 / \sqrt{1+ z^2}$ and expanding
$\operatorname{arsinh}\, z$ into a power series. The Legendre
transform is evaluated separately for odd and even powers in $z$ but
in both cases, we find:
\begin{align}
\frac{1}{\alpha} \int_0^1\frac{\dd z}{\sqrt{\left(\frac{z}{\alpha}
    \right)^2+1}} P_l(z) &= \frac{\sqrt{\pi}}{2 \alpha} {}_3
\tilde{F}_2 \left(1, \frac{1}{2}, \frac{1}{2}; 1 - \frac{l}{2},
\frac{3}{2} + \frac{l}{2}; -\frac{1}{\alpha^2} \right) \, ,
\end{align}
where we have set $\alpha = i a$ and ${}_3 \tilde{F}_2$ is the
regularised hypergeometric function. After some algebra, this reduces
to
\begin{align}
\frac{1}{\alpha} \int_0^1\frac{\dd z}{\sqrt{\left(\frac{z}{\alpha}
    \right)^2+1}} P_l(z) &= \frac{(-1)^{l/2} \alpha^{-l-1}}{2
  \sqrt{\pi} } {}_2 \tilde{F}_1 \left(\frac{1}{2} + \frac{l}{2},
\frac{1}{2} + \frac{l}{2}; \frac{3}{2} + l; -\frac{1}{\alpha^2}
\right) \, .
\end{align}
Taking into account that $\alpha$ is purely imaginary, we finally obtain
the real part:
\begin{align}
\int_a^1\frac{\dd z}{\sqrt{z^2-a^2}} P_l(z) &= \Re \left\{
\int_0^1\frac{\dd z}{\sqrt{z^2-a^2}} P_l(z) \right\} = \sqrt{1-a^2} \;
    {}_2 F_1 \left( \frac{1}{2} - \frac{l}{2}, 1 + \frac{l}{2},
    \frac{3}{2} ,1-a^2 \right) \, .
\end{align}

\bibliographystyle{jhep}
\bibliography{graps}

\end{document}